 \documentclass[acmlarge]{acmart}

\AtBeginDocument{%
  \providecommand\BibTeX{{%
    \normalfont B\kern-0.5em{\scshape i\kern-0.25em b}\kern-0.8em\TeX}}}

\setcopyright{acmcopyright}
\copyrightyear{2018}
\acmYear{2018}
\acmDOI{XXXXXXX.XXXXXXX}

\acmConference[Conference acronym 'XX]{Make sure to enter the correct
  conference title from your rights confirmation emai}{June 03--05,
  2018}{Woodstock, NY}
\acmBooktitle{Woodstock '18: ACM Symposium on Neural Gaze Detection,
 June 03--05, 2018, Woodstock, NY} 
\acmPrice{15.00}
\acmISBN{978-1-4503-XXXX-X/18/06}

\newcommand{\jtrevised}[1]{\textcolor{black}{#1}}
\newcommand{\ryrevised}[1]{\textcolor{black}{#1}}
\newcommand{\clrevised}[1]{\textcolor{black}{#1}}

\newcommand{\lryrevised}[1]{\textcolor{black}{#1}}

\usepackage{longtable}
\usepackage{lscape}
\usepackage{multirow}
\usepackage{appendix}

\begin{document}

\title[Designing Awareness-augmented Telepresence Robot]{TeleAware Robot: Designing Awareness-augmented Telepresence Robot for Remote \jtrevised{Collaborative Locomotion}}


\author{Ruyi Li}
\email{lry_sjtu@sjtu.edu.cn}
\orcid{0009-0005-0710-7171}
\affiliation{%
  \institution{Institute for AI Industry Research, Tsinghua University}
  \city{Beijing}
  \country{China}
}

\author{Yaxin Zhu}
\orcid{0009-0004-8706-4898}
\affiliation{%
  \institution{The Future Laboratory, Tsinghua University}
 \city{Beijing}
  \country{China}
}

\author{Min Liu}
\orcid{0009-0008-8001-3165}
\affiliation{%
  \institution{Institute for AI Industry Research, Tsinghua University}
 \city{Beijing}
  \country{China}
}

\author{Yihang Zeng}
\orcid{0009-0009-5291-0608}
\affiliation{%
  \institution{The Chinese University of Hong Kong, Shenzhen}
 \city{Shenzhen}
  \country{China}
}

\author{Shanning Zhuang}
\orcid{0009-0002-8431-3671}
\affiliation{%
  \institution{Institute for AI Industry Research, Tsinghua University}
 \city{Beijing}
  \country{China}
}

\author{Jiayi Fu}
\orcid{0009-0008-9591-4601}
\affiliation{%
  \institution{Beijing University of Technology}
 \city{Beijing}
  \country{China}
}

\author{Yi Lu}
\orcid{0000-0001-9962-9885}
\affiliation{%
  \institution{Beijing University of Technology}
 \city{Beijing}
  \country{China}
}

\author{Guyue Zhou}
\orcid{0000-0002-3894-9858}
\affiliation{%
  \institution{Institute for AI Industry Research, Tsinghua University}
 \city{Beijing}
  \country{China}
}

\author{Can Liu}
\authornote{Corresponding Author}
\orcid{0000-0003-3267-3317}
\affiliation{%
\institution{School of Creative Media, City University of Hong Kong}
  \city{Hong Kong}
  \country{China}
}
\email{CanLiu@cityu.edu.hk}

\author{Jiangtao Gong}
\authornotemark[1]
\email{gongjiangtao2@gmail.com}
\orcid{0000-0002-4310-1894}
\affiliation{%
  \institution{Institute for AI Industry Research, Tsinghua University}
  \city{Beijing}
  \country{China}
}

\renewcommand{\shortauthors}{Trovato and Tobin, et al.}

\begin{abstract}
\clrevised{Telepresence robots can be used to support users to navigate an environment remotely and share the visiting experience with their social partners. Although such systems allow users to see and hear the remote environment and communicate with their partners via live video feed, this does not provide enough awareness of the environment and their remote partner's activities.
In this paper, we introduce an awareness framework for collaborative locomotion in scenarios of onsite and remote users visiting a place together. From an observational study of small groups of people visiting exhibitions, we derived four design goals for enhancing the environmental and social awareness between social partners,} and developed a set of awareness-enhancing techniques to add to a standard telepresence robot - named TeleAware robot. Through a controlled experiment simulating a guided exhibition visiting task, TeleAware robot showed the ability to lower the workload, facilitate closer social proximity, and improve mutual awareness and social presence compared with the standard one. \clrevised{We discuss the impact of mobility and roles of local and remote users, and provide insights for the future design of awareness-enhancing telepresence robot systems that facilitate collaborative locomotion.}


\end{abstract}

\ccsdesc[500]{Human-centered computing~Interactive systems and tools}
\ccsdesc[500]{Computer systems organization~Embedded systems}
\ccsdesc{Computer systems organization~Robotics}

\keywords{telepresense robot, human-robot interaction, collaborative locomotion, awareness augment}


\begin{teaserfigure}
  \includegraphics[width=\textwidth]{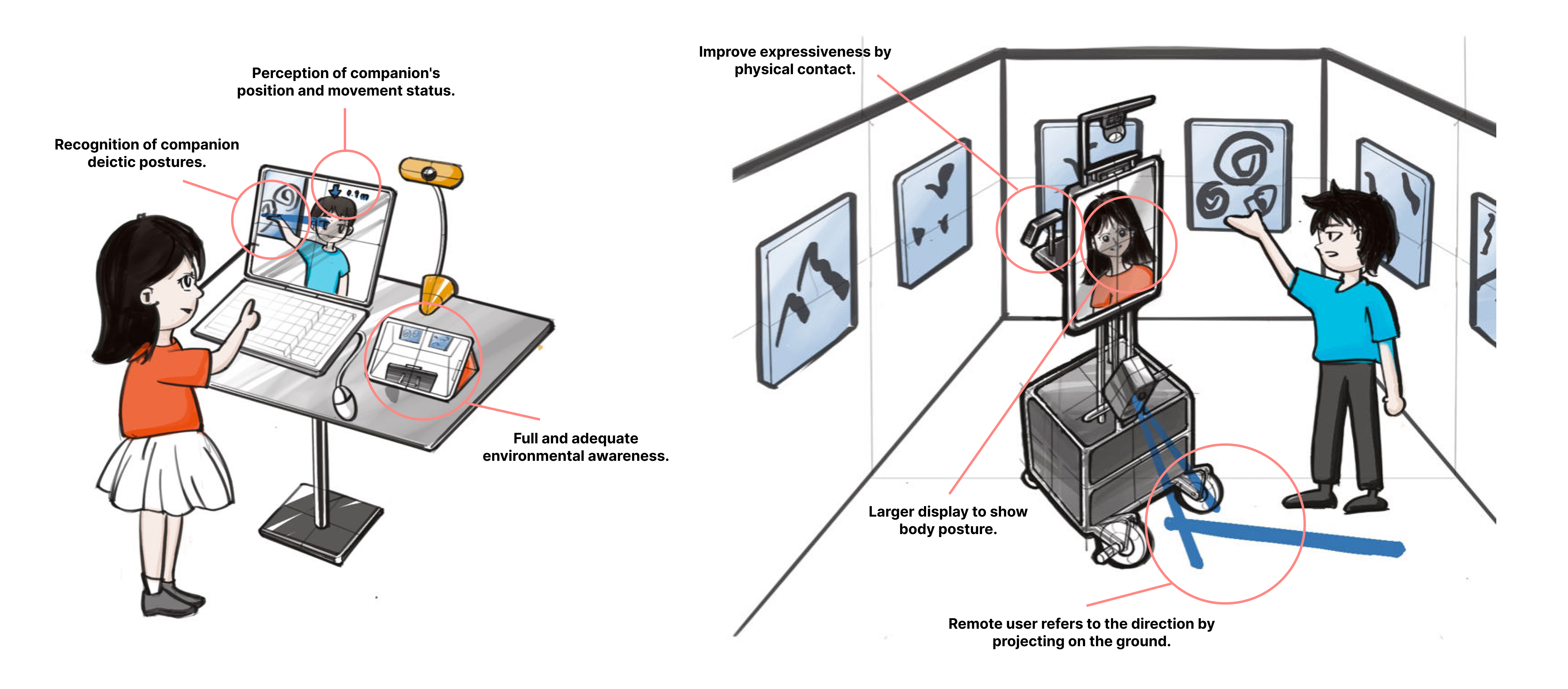}
  \caption{Conceptual Design of TeleAware Robot. In response to remote collaborative locomotion, we have analyzed four design goals based on a "User-Partner-Environment" relationship framework concluded through a series of observational studies. Building upon these goals, we have designed novel features to enhance the telepresence robot, enabling the TeleAware Robot to improve mutual awareness to better support collaborative locomotion.}
  \Description{.}
  \label{fig:teaser}
\end{teaserfigure}

\received{20 February 2007}
\received[revised]{12 March 2009}
\received[accepted]{5 June 2009}

\maketitle

\section{Introduction}
Remote collaboration occurs more and more frequently in people's daily lives today for various work, study, and leisure activities. Sometimes one or more of the parties are moving within a physical space or in transit~\cite{alem2011developing,pejoska2017mobile,wu2023mr}. In a physical work context, remote collaborative locomotion is widely used in many real-world scenarios, such as academic conference attendance~\cite{neustaedter2016beam}, and joint shop exploration~\cite{yang2018shopping}.
\jtrevised{When people walk together with others in the same location, they have a strong awareness of the environment and their social partners through subconscious locomotion 
~\cite{rasch2023going}. Typically, 
social partners engage in actions such as moving, guiding, roaming, navigating, and sharing points of interest in an open space. Through these actions, people become aware of each others' status and intentions with environmental references. Such subtle information supports their collaborative navigation and constructs a natural shared visiting experience.}

However, \clrevised{collaborative locomotion with a telepresence robot is known to be challenging~\cite{young2020mobileportation,yang2018shopping,neustaedter2016beam}. Previous work found it hard for the remotely controlled robot to follow the onsite user. 
While this experiment showed significant benefits 
for researchers unable to attend conferences remotely, it also revealed the limitations of telepresence robots in such collaborative locomotion tasks. For instance, remote operators may have insufficient awareness of spatial proximity to the environment or partners, and on-site social interactions.}


Enhancing remote collaboration has long captured the attention of researchers in the fields of human-computer interaction and ubiquitous computing~\cite{neustaedter2016beam,guo2019blocks,young2020mobileportation}. 
In recent years, there have been numerous system innovations and empirical studies concerning telepresence robots. For example, in the context of collaborative design~\cite{sakashita2023remotion}, collaborative puzzle-solving activities~\cite{sakashita2022remotecode}, collaborative shopping scenarios~\cite{yang2018shopping}, and by incorporating augmented reality~\cite{jones2021belonging} or providing more non-verbal information~\cite{stahl2018social,wang2023gesture}, efforts have been made to enhance their cooperative performance and collaborative experience. 
\jtrevised{However, due to the limited awareness of remote users in collaborative locomotion tasks involving telepresence robots, most previous studies have not treated telepresence robots as equal entities to local collaborators. Most telepresence robots are relegated to the role of followers, rather than being capable of acting as leaders. This limits the participation of remote users and sacrifices their experience. }

In this work, our primary goals are to understand how to enhance the social and environmental awareness between local and remote users and to design a set of awareness-enhancing features for telepresence robots to support users' remote collaborative locomotion experience.  
Our initial inspiration came from a series of observational studies of people visiting exhibitions in-person in pairs. This is a typical shared experience with much collaborative locomotion navigating a place. ~\lryrevised{The observational research comprises the pilot study and the supplementary study, with the latter conducted after controlled experiments to validate findings obtained from the pilot study. }
Based on the naturally occurring patterns we observed in the studies, we constructed a design space for user-partner-environment awareness and derived four design goals for enhancing awareness of telepresence robots. We implemented example features for each design goal and added them to a standard telepresence robot, which we named as "TeleAware Robot." Through controlled user experiments, we compared the performance, navigation patterns, awareness, and subjective feedback of the TeleAware Robot against a standard telepresence robot in remote collaborative tasks. Finally, we summarized the experimental results and provided implications for the future design of telepresence robots in collaborative locomotion scenarios.

Our contributions can be summarized as follows:
\jtrevised{i) An awareness framework derived from an observational study of natural human behavior for collaborative locomotion in an example scenario of colocated dyads visiting exhibitions together. 
ii) Based on this framework, we introduced four design goals to enhance the awareness of telepresence robots in supporting collaborative locomotion and designed a set of awareness-enhancing features to reach these goals. 
iii) We built TeleAware Robot by adding these features to a standard telepresence robot and tested its effectiveness on enhancing awareness and improving collaborative locomotion in comparison with a standard robot as the baseline. 
iv) We contribute insights from an in-depth analysis and discussion on the effects of the awareness-enhancing features on users' roles, interaction and presence, informing the future design of telepresence for collaboration.}

\section{Related Work}
\subsection{Collaborative Locomotion}
~\ryrevised{When groups engage in collaborative activities in open spaces, collaborative locomotion is often observed. Humans frequently move together in formations ranging from small to large groups. These natural systems exhibit ordered and coherent patterns of motion, considered to be self-organized \cite{couzin2003self,helbing2001self}. As we walk with others in the physical world, we unconsciously adjust our stride frequency and length to maintain close proximity to our partners \cite{rasch2023going}. Synchronized movement is associated with increased liking \cite{hove2009s}, higher cooperation \cite{reddish2013let}, and trust \cite{launay2013synchronization} among people. Previous research has identified leadership as a significant feature in collaborative locomotion \cite{king2010follow,zhang2023follower}, defined by Krause et al. \cite{krause2000leadership}as the initiation of a new movement direction by one or more individuals, subsequently followed by other group members. Leadership and following can be passive processes, where individuals adopt the roles of leaders and followers, organizing themselves without explicit communication or understanding of each other's roles \cite{couzin2003self}. }

\jtrevised{Previous research has conducted studies on collaborative locomotion in both virtual environments and the field of telepresence robots.}
~\ryrevised{In virtual reality environments, numerous studies have explored group collaborative locomotion. Weissker et al. \cite{weissker2019multi}utilized an analogy similar to a dual driver/passenger scenario in their Multi-Ray-Jumping game, where one player is responsible for selecting the teleportation destination for both individuals. In a subsequent paper by Weissker et al. \cite{weissker2020getting}, this mechanism was refined, allowing passengers to choose their positions before engaging in group teleportation. Weissker and Froehlich \cite{weissker2021group} further developed this concept to accommodate more than just two individuals, enabling groups of up to ten participants to teleport in guided tours with configurable position settings. Rasch et al. \cite{rasch2023going} studied three visualization effects in a leader-follower experimental setup, finding that shared visualizations support understanding of movement intentions, increasing group cohesion while maintaining individual movement freedom. In the domain of telepresence robots, remote users manipulating robots often act as followers in collaborative locomotion. Previous research has enhanced the navigational capabilities of telepresence robots to follow specific individuals in the environment, thus enhancing the collaborative locomotion capabilities of both parties \cite{trahanias2000tourbot,burgard2003tele}. Studies indicate that remote users relying on telepresence robots often heavily depend on local users, typically acting as followers \cite{yang2018shopping}. Remote users assume a leading role only when they have a strong reason to do so.}


\jtrevised{Despite prior studies on collaborative locomotion, research focusing on awareness in collaborative locomotion remains scarce. Telepresence robots, used as communication intermediaries, lack systematic design solutions to enhance awareness. The field also overlooks discussions on dominance in coordinated movements and equal leadership empowerment. Telepresence robots are often limited to follower roles, impeding their dominance and naturalness in remote collaboration. Therefore, our work targets these specific challenges in collaborative locomotion tasks involving telepresence robots.}

\subsection{Telepresence Robot}
Telepresence robots are remotely operated robotic devices designed to provide individuals with a physical sense of presence including remote control and video communication \cite{rosenberg2019human}. Remote control allows remote users to move the robot and change its direction through computers, mobile devices, and other user interfaces. Telepresence robots are applied in various fields including business \cite{yang2018shopping}, healthcare \cite{ruiz2021mental}, education \cite{botev2020immersive}, and personal use \cite{adalgeirsson2010mebot}, providing effective means to bridge geographical distances and enhance remote communication, collaboration, and exploration. 

Current research utilizing telepresence robots for remote collaboration has explored various collaboration scenarios, such as remote meetings, design collaboration \cite{sakashita2023remotion}, and puzzle-solving activities \cite{sakashita2022remotecode}. For instance, Sakashita et al. \cite{sakashita2023remotion} have investigated robot design for remote design collaboration in open spaces. Yang et al. \cite{yang2018shopping} have utilized telepresence robots for remote collaborative shopping activities, observing user behavior in such scenarios. Moreover, a considerable amount of research focuses on the design of physical embodiment and its impact on interpersonal communication \cite{ching2016design,rae2013body}. For example, Onishi and colleagues have proposed a method in video conferences to embody remote partners' body parts, synchronizing robot arm movements with partners’ actions to reduce perceived distance \cite{onishi2016embodiment,bazzano2017comparing}. As robot embodiment provides more non-verbal cues, such as gestures \cite{stahl2018social,wang2023gesture}, postures \cite{hasegawa2014telepresence}, motions \cite{li2019communicating}, gaze \cite{prajod2023gaze}, and facial expressions \cite{yonezawa2014representation}, it enhances the richness of communication.

However, there are still areas to be explored in using telepresence robots for remote collaboration. Firstly, while some studies have focused on remote collaboration scenarios, these scenarios often revolve around relatively fixed work platforms or spaces. Some studies involving open spaces still center around physical artifacts as the core of mobility \cite{sakashita2023remotion}. Thus, there is less focus on collaboration scenarios involving collaborative locomotion in open spaces. Secondly, many efforts are being made to provide physical functions of robots to enhance remote users' ability to express non-verbal cues, but less attention is paid to enhancing local users' state information. Most systems in these studies have explored remote users' control methods, with feedback perceived by them primarily expressed through video captured by cameras on displays \cite{matsuda2016scalablebody}. However, increased mobility brings many variables to collaborative tasks, and simple visual capture has limitations in providing remote users with cognitive awareness of their local partners' states. 

Therefore, in this paper, we aim to explore design solutions based on standard telepresence robots for enhancing task performance and user experience in collaborative locomotion activities. 

\subsection{Awareness Augmentation for Collaboration}
~\ryrevised{Awareness is a concept that encompasses the processes of knowing, perceiving, and being cognizant of events \cite{smith2017self}.} ~\ryrevised{From the perspectives of psychology and neuroscience, awareness and attention are closely related; however, awareness permits the simultaneous experience of multiple stimuli, whereas attention is a selective and in-depth processing mechanism~\cite{lamme2003visual,gong2021holoboard}.} In collaborative processes, due to the numerous factors related to others that need to be understood as a context for own activity, enhancing awareness of multiple relevant factors has always been a focal point of Computer Supported Cooperative Work research~\cite{dourish1992awareness,gong2023side}.
Collaborative awareness involves team members having knowledge and understanding of each other's real-time actions, intentions, and states within a collaborative task. Additionally, many researchers use additional terms to specify their focus on collaborative awareness, including task awareness \cite{fransen2011mediating}, situational awareness \cite{gao2023agent}, social awareness \cite{lambropoulos2012supporting}, spatial awareness \cite{fink2022re}, peripheral awareness \cite{sakashita2022remotecode}, and other concepts. ~\ryrevised{In environments characterized by high mobility within remote collaboration, traditional audio-visual communication media, and desktop-based collaborative tools may have limitations in supporting face-to-face interactions and enhancing the experience of collaborative awareness \cite{teo2019mixed,wu2023mr}.} A substantial amount of work has focused on improving various elements of remote collaboration. For example, the CoVAR system attempts to create novel forms of remote collaboration using augmented reality and virtual reality, enhancing the expression of natural communication cues \cite{piumsomboon2017covar,lu2022classification}. Teo et al., through a combination of 360-degree video and 3D virtual reconstruction, have enhanced situational awareness and spatial understanding in remote collaboration \cite{teo2019mixed}.

However, despite substantial research efforts addressing some aspects of remote collaboration, there has been relatively limited research specifically targeting collaborative locomotion activities. The high degree of mobility in this scenario introduces more complex requirements for collaborative awareness. Currently, there is a lack of systematic design solutions or enhancement strategies proposed for meeting these requirements when utilizing telepresence robots as communication intermediaries. 

Therefore, our research will investigate whether telepresence robots can be systematically awareness augmented and how these enhancements affect task performance and collaborative experience through empirical studies.

\section{Observational Study}

To inspire the design of a telepresence robot in ~\ryrevised{collaborative locomotion activities}, we conducted an observational study. ~\ryrevised{We chose the "dyad" as the minimal unit of collaboration, using the typical scenario of "collaborative exhibition viewing" as our case study.} ~\ryrevised{As our goal was to understand the specific awareness needs of collaborators in collaborative locomotion, we focused on observing the typical behaviors and collaborative methods of viewers in pairs to better support the design of telepresence robots.}

~\lryrevised{Our observational study comprises two components: the pilot study and the supplementary study. The pilot study is conducted through the observation of a small sample for qualitative analysis, whereas the supplementary study is carried out after the completion of design and controlled experiments, aiming to validate the conclusions of the pilot study through quantitative analysis methods. For ease of discussion, both the pilot study and supplementary research will be jointly elucidated in this chapter.}

~\lryrevised{In our pilot study, observations were conducted at two exhibitions with different themes: a graduation design exhibition and a product exhibition.  In selecting these exhibitions, we particularly considered the factor of visitor flow. By observing situations with varying degrees of crowding, we aimed to explore the commonalities in dual-user viewing behavior and assess the potential impact of different visitor flows on observation behaviors. The graduation design exhibition had a relatively low visitor flow, with most areas appearing quite vacant. This exhibition primarily featured display panels with rich graphical and textual materials, presenting content at varying heights. In contrast, the product exhibition had noticeably higher visitor traffic, with most exhibit areas experiencing dense visitor gatherings, creating a relatively crowded atmosphere.}

~\lryrevised{In our supplementary study, we expanded our observational data with two exhibitions of different themes: a natural history exhibition and a cultural relics exhibition. The visitor flow at these exhibitions was overall high, with the majority of exhibition halls exhibiting a relatively crowded atmosphere, whereas a few halls were more vacant. These two exhibitions primarily showcased physical artifacts, supplemented by graphics, videos, and other presentation methods.} Throughout the observational study, we employed a non-interventional approach to allow participants to interact naturally and freely.


\begin{figure}[ht]
    \centering
    \includegraphics[width=0.6\linewidth]{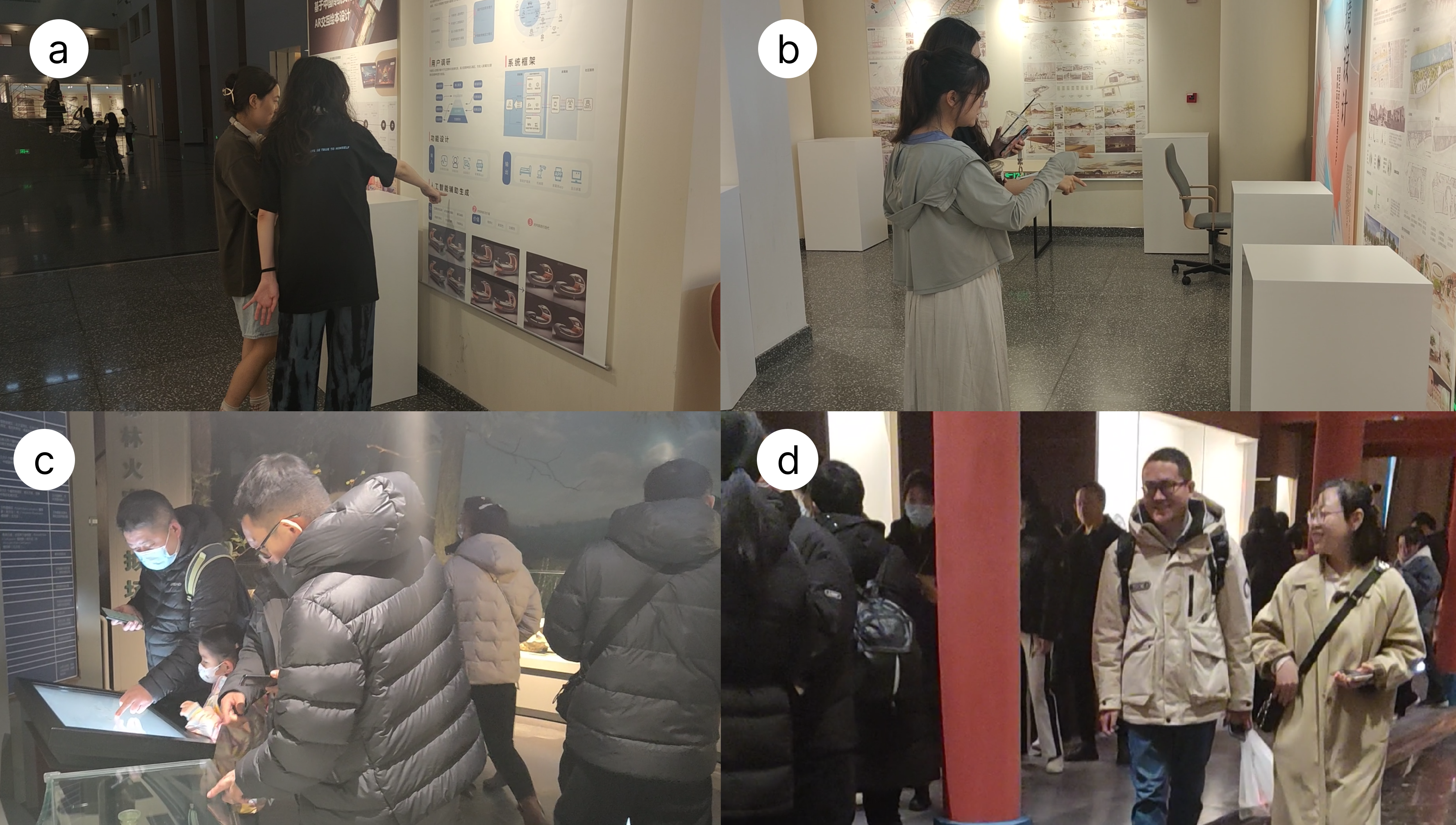}
    \caption{a) and b) are photos from the observational study in the exhibition with low visitor flow. c) and d) are photos from the exhibitions with higher visitor flow.}
    \label{fig:observation}
\end{figure}

\subsection{Participants}
~\lryrevised{Our study involved a total of 38 participants (19 pairs). Of these, six pairs were recruited for the pilot study, with an age range from 23 to 34 years (M = 26.08, SD = 3.84, comprising three males and nine females). Thirteen pairs were recruited for the supplementary study, with an age range from 16 to 65 years (M = 28.77, SD = 10.69, comprising 15 males and 11 females). To ensure that the observational study results comprehensively covered the range of behaviors that might occur during dual-user exhibition viewing, different gender combinations were considered. Among these, seven pairs of visitors were female-female combinations, six pairs were male-female combinations, and the remaining six pairs were male-male combinations. }
\subsection{Procedure}
In our observational study, we sought and recruited pairs of exhibition visitors at the venue. ~\ryrevised{Each pair was introduced to our objective, namely, an observational study focused on dual-user viewing behavior. The visitors were informed that we would be observing them from a distance and recording their activities via video to minimize disturbance to their exhibition experience. All participants consented to the collection of video data, and each pair of viewers was observed for ten minutes. We documented the entire process of dual-user viewing through video recordings, which were used for subsequent analysis.}

\subsection{Data Analysis}

\begin{table}[]
\caption{The video coding codebook based on the "User-Partner-Environment" framework.}
\begin{tabular}{lllll}
\hline
\textbf{Relationship}                                                             & \textbf{\begin{tabular}[c]{@{}l@{}}Behavior\\ Code\end{tabular}} & \textbf{\begin{tabular}[c]{@{}l@{}}Behavior\\ Category\end{tabular}}                         & \textbf{Operational Definition}                                                                                                                                                           & \textbf{Subcategories}                                                                                           \\ \hline
\textbf{\begin{tabular}[c]{@{}l@{}}user-\\ environment\end{tabular}}              & \multirow{2}{*}{ES}                                              & \multirow{2}{*}{\begin{tabular}[c]{@{}l@{}}Visual\\ Perception\end{tabular}}                 & \multirow{2}{*}{\begin{tabular}[c]{@{}l@{}}Viewers observe\\ information\\ in the environment.\end{tabular}}                                                                              & \begin{tabular}[c]{@{}l@{}}ES1: Viewer scans\\ the environment\\ (no specific target).\end{tabular}              \\ \cline{5-5} 
\textbf{}                                                                         &                                                                  &                                                                                              &                                                                                                                                                                                           & \begin{tabular}[c]{@{}l@{}}ES2: Viewer looks at\\ exhibits or \\ other people \\ (specific target).\end{tabular} \\ \cline{2-5} 
\textbf{}                                                                         & AP                                                               & \begin{tabular}[c]{@{}l@{}}Auditory\\ Perception\end{tabular}                                & \begin{tabular}[c]{@{}l@{}}Viewers pay attention\\ to other sounds\\ in the environment.\end{tabular}                                                                                     &                                                                                                                  \\ \cline{2-5} 
\textbf{}                                                                         & CO                                                               & \begin{tabular}[c]{@{}l@{}}Communication \\ with Others\\ in the \\ Environment\end{tabular} & \begin{tabular}[c]{@{}l@{}}Verbal communication\\ with people \\ other than the companion \\ in the environment.\end{tabular}                                                             &                                                                                                                  \\ \hline
\textbf{\begin{tabular}[c]{@{}l@{}}user-\\ partner\end{tabular}}                  & \multirow{2}{*}{VS}                                              & \multirow{2}{*}{\begin{tabular}[c]{@{}l@{}}Visual\\ Perception\end{tabular}}                 & \multirow{2}{*}{\begin{tabular}[c]{@{}l@{}}The viewer's gaze is \\ directed towards \\ or searching for\\ the partner.\end{tabular}}                                                      & \begin{tabular}[c]{@{}l@{}}VS1: directly\\ towards the partner.\end{tabular}                                     \\ \cline{5-5} 
\textbf{}                                                                         &                                                                  &                                                                                              &                                                                                                                                                                                           & \begin{tabular}[c]{@{}l@{}}VS2: search\\ for the partner.\end{tabular}                                           \\
                                                                                  &                                                                  &                                                                                              &                                                                                                                                                                                           &                                                                                                                  \\ \cline{2-5} 
\textbf{}                                                                         & VC                                                               & \begin{tabular}[c]{@{}l@{}}Verbal\\ Communication\end{tabular}                               & \begin{tabular}[c]{@{}l@{}}Viewers engage in verbal\\ communication\\ with their partner.\end{tabular}                                                                                    &                                                                                                                  \\ \cline{2-5} 
\textbf{}                                                                         & \multirow{3}{*}{PC}                                              & \multirow{3}{*}{\begin{tabular}[c]{@{}l@{}}Physical\\ Contact\end{tabular}}                  & \multirow{3}{*}{\begin{tabular}[c]{@{}l@{}}Physical contact occurs\\ between the viewer\\ and the partner.\end{tabular}}                                                                  & \begin{tabular}[c]{@{}l@{}}PC1: touch\\ shoulder or arm.\end{tabular}                                            \\ \cline{5-5} 
\textbf{}                                                                         &                                                                  &                                                                                              &                                                                                                                                                                                           & PC2: walk arm-in-arm.                                                                                            \\ \cline{5-5} 
\textbf{}                                                                         &                                                                  &                                                                                              &                                                                                                                                                                                           & PC3: hold hands.                                                                                                 \\ \cline{2-5} 
\textbf{}                                                                         & \multirow{2}{*}{WG}                                              & \multirow{2}{*}{\begin{tabular}[c]{@{}l@{}}Waving\\ Gesture\end{tabular}}                    & \multirow{2}{*}{\begin{tabular}[c]{@{}l@{}}Viewer waves to\\ the companion.\end{tabular}}                                                                                                 & \multirow{2}{*}{}                                                                                                \\
\textbf{}                                                                         &                                                                  &                                                                                              &                                                                                                                                                                                           &                                                                                                                  \\ \hline
\textbf{\begin{tabular}[c]{@{}l@{}}user-\\ (partner\\ -environment)\end{tabular}} & ET                                                               & \begin{tabular}[c]{@{}l@{}}Eye\\ Tracking\end{tabular}                                       & \begin{tabular}[c]{@{}l@{}}One viewer looks at \\ an exhibit, and the \\ other pne follows to \\ the same exhibit.\end{tabular}                                                           &                                                                                                                  \\ \cline{2-5} 
\textbf{}                                                                         & GR                                                               & \begin{tabular}[c]{@{}l@{}}Gestural\\ Reference\end{tabular}                                 & \begin{tabular}[c]{@{}l@{}}Viewer uses gestures\\ to refer to\\ the environment\\ or exhibits.\end{tabular}                                                                               &                                                                                                                  \\ \cline{2-5} 
\textbf{}                                                                         & \multirow{6}{*}{DO}                                              & \multirow{6}{*}{Leadership}                                                                  & \multirow{6}{*}{\begin{tabular}[c]{@{}l@{}}Two viewers act as\\ the leader or follower, \\ with the follower\\ moving and communicating\\ in accordance \\ with the leader.\end{tabular}} & DOP1:P1 has leadership.                                                                                          \\
                                                                                  &                                                                  &                                                                                              &                                                                                                                                                                                           &                                                                                                                  \\ \cline{5-5} 
\textbf{}                                                                         &                                                                  &                                                                                              &                                                                                                                                                                                           & DOP2:P2 has leadership.                                                                                          \\
                                                                                  &                                                                  &                                                                                              &                                                                                                                                                                                           &                                                                                                                  \\ \cline{5-5} 
\textbf{}                                                                         &                                                                  &                                                                                              &                                                                                                                                                                                           & \begin{tabular}[c]{@{}l@{}}NDO: Leadership is\\ indiscernible.\end{tabular}                                      \\
                                                                                  &                                                                  &                                                                                              &                                                                                                                                                                                           &                                                                                                                  \\ \hline
\end{tabular}
\label{Tab:CodeBook}
\end{table}

In the pilot study, we conducted a qualitative analysis of the video data. We were inspired by Kuflik and Dim's categorization of dual-user exhibition viewing behaviors~\cite{dim2014automatic}, based on the degree of social synchronization between users and their partners and the level of attention users give to exhibits in their environment. During the qualitative analysis phase, we preliminarily identified an interaction framework centered around the "User-Partner-Environment" triad in dual-user collaborative viewing scenarios. In the supplementary study, to thoroughly analyze the behaviors and awareness needs in collaborative locomotion, we conducted a video coding process on the recorded video data. In developing our codebook, we based our analysis on the interaction framework derived from the pilot study. Accordingly, we analyzed behaviors that needed coding concerning User-Environment, User-Partner, and User-(Partner-Environment) interactions and developed a codebook based on these. The codebook is shown in Table~\ref{Tab:CodeBook}.

~\ryrevised{The video coding was carried out using the YueLiu~
\footnote{\url{https://yueliu.cn/}} video annotation platform. When a behavior requiring coding was observed among viewers, we marked the corresponding behavior code. After completing the annotations, we performed counts, or duration statistics for the behavior codes. Two members were involved in the video coding. Before starting the formal coding, both parties familiarized themselves with and discussed the behaviors involved in the codebook. We selected a video from the data that broadly covered all behaviors in the codebook for a test coding, with each coding member independently completing the coding for this video. After coding, the members compared and analyzed their results, discussing inconsistencies in the coding and establishing new standards. Furthermore, both parties discussed newly observed codable behaviors and iteratively updated the codebook. After two rounds of iteration, the consistency in coding between the two coders reached a percentage of 95\%.}

\subsection{Findings}\label{Sec:Findings}
~\lryrevised{In our observational study, we collected a total of 170 minutes of video data. Of this, 40 minutes were from the pilot study and 130 minutes from the supplementary study. During the quantitative analysis process, video data for three pairs of participants from the pilot study were too scant for subsequent quantitative coding. Thirteen minutes of data from the supplementary study were deemed invalid because the viewers were separated for extended periods, rendering effective analysis of collaborative behavior unfeasible. Therefore, the total duration of valid data for quantitative analysis amounted to 147 minutes, involving 16 pairs of participants (three pairs from the pilot study and thirteen pairs from the supplementary study), with an age range from 16 to 65 years (M = 27.72 years, SD = 9.88, comprising 15 males and 17 females).}
\ryrevised{We coded the video data according to the final version of our codebook and summarized the results. A few behaviors listed in the codebook were scarcely observed; among the 16 participant pairs, there was only one instance of a waving gesture and one instance of communication with other people in the environment. Thus, we considered these behaviors as not constituting the main activities in collaborative locomotion. Based on our observational outcomes, we derived the "User-Partner-Environment" core framework and analyzed the awareness needs and corresponding behaviors within this framework. We will report our findings according to the three relationships: "User-Environment," "User-Partner," and "User-(Partner-Environment)."}

\begin{figure}[ht]
    \centering
    \includegraphics[width=0.8\linewidth]{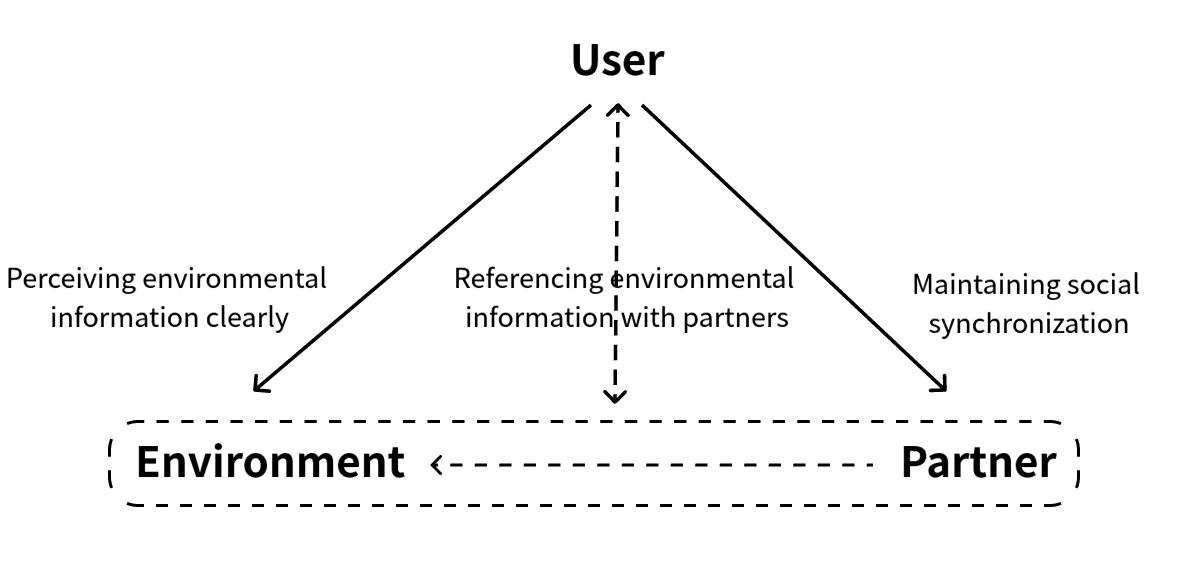}
    \caption{\ryrevised{In collaborative locomotion, the "User-Partner-Environment" framework diagram primarily encompasses three relationships: "User-Environment," "User-Partner," and "User-(Partner-Environment)." The User-Environment relationship is founded on spatial awareness, with the objective of clearly perceiving the environment. The User-Partner relationship is grounded in social awareness, aiming to maintain social synchronization. The User-(Partner-Environment) relationship, based on situational awareness, signifies partners' aspiration to form a joint reference to environmental information with each other. The two solid lines represent the relationships between the user and the environment, and between the user and the partner, respectively. Meanwhile, the dashed line, in conjunction with the dashed box, collectively signifies the User-(Partner-Environment) relationship. Arrows indicate the user's demands towards the respective entities, highlighting the directional nature of these requirements within the framework. Bidirectional arrows represent the mutual need for expression and understanding between the user and partners in the User-(Partner-Environment) relationship.}}
    \label{fig:framework}
\end{figure}

\subsubsection{User-Environment}
~\ryrevised{Our observational data indicates that in dual-user collaborative exhibition viewing scenarios, viewers exhibit a strong sense of spatial awareness. Spatial awareness refers to the user's awareness of their relationship with the surrounding environment and themselves. People are situated in environments and process spatial information about those environments to understand their position within them. Simultaneously, people interact with and think about their surroundings, and their knowledge and perceptions of the environment, in turn, guide their behaviors in an adaptive manner \cite{ishikawa2021spatial}. Therefore, perceiving environmental information is fundamental to spatial awareness and behavior.}

~\ryrevised{In our observational study, we validated the importance of the visual perceptual channel for users in processing spatial information. Viewers spent most of their time observing the environment. We conducted a statistical analysis of the time distribution, finding that the time spent looking around the environment without a specific target accounted for 16.12\%, while the time spent looking at exhibits or other people in the environment accounted for 70.05\%. A clear understanding of the environment enabled viewers to move autonomously and plan their exhibition routes. Simultaneously, we observed that users could acquire environmental information through the auditory channel, but the number of instances where they were attracted by environmental sounds was relatively few. In our observations, six groups of participants were drawn to sounds in the environment, with two groups noticing sudden environmental noises in the spacious exhibition and the remaining four groups noticing the sound from video displays in crowded environments. We found that compared to the auditory channel, the visual perceptual channel played a more dominant role in users' understanding of the environment. Especially in crowded environments, higher foot traffic resulted in noisier surroundings, and viewers tended to focus on verbal communication with each other, rather than being easily distracted by sudden sounds in the environment. In such scenarios, obtaining spatial information visually became the core method, facilitating users' cognition of the overall environment.}

~\ryrevised{In remote collaborative conditions, the remote user's ability to perceive the local environment visually is more limited. Although remote communication technologies can provide high-quality video transmission, the remote user's perspective is often restricted by the position and angle of the camera. This means they cannot obtain a comprehensive view of the environment. The results of our observational study inspire us to enhance the environmental visibility of local spaces for remote users in the design of telepresence robots, to improve spatial awareness. Therefore, we formulate our Design Goal 1:}

\begin{quote}
    \textbf{Design Goal 1: Enhance the environmental visibility of local spaces for remote users to improve spatial awareness.}
\end{quote}

\subsubsection{User-Partner}
~\ryrevised{Our observational data indicates that in dual-user collaborative exhibition viewing scenarios, viewers exhibit a strong sense of social awareness. Social awareness refers to the awareness of the social situation of collaborative members \cite{tollmar1996supporting}, i.e., being aware of what they are doing, what they are interested in, and their emotional states, among other aspects. A high level of social awareness fosters efforts by both collaborators to achieve a high degree of social synchronization, meaning members maintain good coordination and consistency in behaviors and cognitive states. We observed that pairs could naturally follow each other in sequence or walk side by side, maintaining synchronous locomotion patterns. Simultaneously, synchronous social attention enabled viewers to engage in active communication. In order to maintain this synchrony, members continually collected information about their partners and took corresponding actions.}
\paragraph{Visual Perception}
~\ryrevised{In scenarios without physical interaction, visual perception is a key behavioral method for acquiring information about partners. Looking towards a partner directly provides information about their location and status, with this behavior occurring at an average frequency of 1.6 times per minute. We also found that the visitor flow of the exhibition influenced the behavior of viewers visually searching for their companions. Among the 16 participating pairs, four exhibited behavior of visually searching for a lost partner, and these four pairs were all recruited from the two exhibitions with high visitor flow. Due to the high degree of crowding, the 16th pair of participants even displayed searching behavior six times during the 10-minute observation period. When participants regained perception of their partner's location and status, they usually moved towards them to form new social synchronization. The observation results suggest that high visitor flow can induce new visual searching behaviors, but the flow itself does not affect the core purpose of visual perception of the partner, which is to obtain information about their location and status in order to maintain social synchronization. 
Additionally, visual perception of a partner can assist exhibition viewers in maintaining and transitioning leadership roles in collaborative locomotion activities. During observations, leaders looked towards their companions to confirm if they were closely following, while followers used visual cues to coordinate how to follow the leader. In situations where both parties had equal leadership roles, a new leader emerged when one party visually detected a sudden change in their partner's movement direction. This underscores the importance of visual perception of partners in collaborative locomotion activities.}

~\ryrevised{Under remote collaboration conditions, due to the physical presence of the telepresence robot, local users often have a very direct understanding of the remote user's location and status \cite{yang2018shopping}. However, remote users rely on cameras to capture the local environment, and due to device factors, as well as instances when the local user is in the robot's blind spot or is obscured, remote users may find it more challenging to perceive the state of their collaborators. The results of our observational study inspire us to support remote users in perceiving the location and status of local users in the design of telepresence robots, in order to improve social awareness. Therefore, we formulate our Design Goal 2:}
\begin{quote}
    \textbf{Design Goal 2: Support remote users in perceiving the location and status of
local users to improve social awareness.}
\end{quote}

\paragraph{Physical Contact}
~\ryrevised{In scenarios involving physical interaction, physical contact is also a common behavior for perceiving the state of a partner. When engaging in physical interactions, the force and torque perceived by partners through tactile communication convey valuable information, determining each participant's role in the task, allowing pairs to collaborate better \cite{evrard2009homotopy}\cite{reed2008physical}. Physical contact allows social intentions such as intimacy, support, reminders, and connection to be effectively conveyed non-verbally, facilitating good social synchronization under social awareness.}

~\ryrevised{In our observational study, nine out of 16 pairs of participants engaged in physical contact interactions, with five pairs being female-female combinations and three pairs male-female combinations. We did not observe a clear impact of visitor flow on physical contact interactions, but we believe that gender and relationship have a more significant influence on the frequency of these interactions. Female-female participant pairs were more willing to use physical contact to convey their social intentions and emotional states. The frequency of male-female physical contact interactions depended on the nature of their relationship, with couples in intimate relationships naturally engaging in physical interactions. The only male-male pair we observed engaging in physical contact had a familial relationship. Additionally, different types of physical contact interactions were observed in the study. Touching a partner's shoulder or arm with a hand can arouse their social attention and convey one's social intentions. Seven pairs of participants touched their partner's shoulder or arm, with an average frequency of 3.3 times per pair over a 10-minute period. Linking arms and holding hands helped viewers maintain intimacy and good internal coordination. We observed five pairs of participants moving while linking arms, with an average frequency of 3.8 times per pair over 10 minutes. Two pairs engaged in hand-holding, with one couple holding hands throughout the entire exhibition viewing. Furthermore, physical contact supported the transition of leadership roles in collaborative locomotion. During observations, we found that when the attention of exhibition viewers and their partners were misaligned, the viewers would tap their partner's shoulder to direct their partner's attention to follow their own. In this non-verbal cue of physical contact communication, the roles of follower and leader were effectively transitioned.}

~\ryrevised{However, for remote and local users under telepresence robot conditions, natural physical contact and feedback are often lacking. This means that conveying social intentions and transitioning leadership roles are difficult to achieve through many intimate and engaging interactive actions. Previous studies have shown that social and tangible combinations are supportive embodied
interaction paradigms for conveying non-verbal cues like physical contact \cite{saadatian2013personalizable}. The results of our observational study inspire us to enhance embodied interaction between local and remote users in the design of telepresence robots, in order to improve social awareness. Therefore, we formulate our Design Goal 3:}
\begin{quote}
    \textbf{Design Goal 3: Enhance embodied interaction between local and remote users to improve social awareness.}
\end{quote}

\subsubsection{User-(Partner-Environment)}
~\ryrevised{Our observational data shows that in dual-user collaborative exhibition viewing scenarios, viewers exhibit a strong sense of situational awareness. Situational awareness involves "understanding what is happening around," where both collaborators are aware of events in their environment and have a mutual understanding of the information \cite{kulyk2008situational}. This awareness relies on joint references to environmental information by users and their partners. We found that following a partner's gaze and observing their gesture indications effectively enable collaborators to form a joint reference to the same focus. These behaviors were universally observed in both spacious and crowded exhibition settings, with gaze following occurring at a frequency of 1.5 times per minute and gesture indication at 2.5 times per minute. Concurrently, we believe that this reference is closely related to the dynamics of dominance within the collaborative group. In our observational study, among 16 participant pairs, 12 demonstrated dynamic role switching between leaders and followers. Both leaders and followers actively used gesture indications to express their environmental references, facilitating dynamic transfers of leadership. Followers, in comparison to leaders, tended to follow the leaders' gaze more to understand the leaders' references, creating a following of the focal point. In the effective time coding of 28 participants acting as leaders, 21 followers were observed to follow the gaze more frequently than the leaders. Specifically, followers were observed to follow the gaze 159 times, while leaders were observed only 59 times.}

~\ryrevised{In remote collaboration, previous research suggests that local users, due to their stronger physical capabilities, often take on the role of leaders in the collaboration process. This results in users operating telepresence robots being highly dependent on local users and perpetually assuming the role of followers, which contrasts with the dynamic power transfer seen in real-world scenarios \cite{yang2018shopping}. In certain cases, specific tasks may require remote users to take on a greater leadership role. Past studies have enhanced the indicative capabilities of remote users to increase their dominance, such as remote experts using projected light points to convey their assistance opinions to local workers \cite{machino2006remote}. Our observations indicate that the transfer of dominance is highly flexible, with both parties possessing leadership equally. The results of our observational study inspire the design of telepresence robots to support both local and remote users in jointly referencing environmental information, developing dominance capabilities, and enhancing situational awareness.}

\begin{quote}    
\textbf{Design Goal 4: Support joint referencing of environmental information by both local and remote users to improve situational awareness.}
\end{quote}

\section{DESIGN}

\subsection{Designing TeleAware Robot Based on Four Design Goals}
~\ryrevised{As we discussed in Section~\ref{Sec:Findings}, we derived a framework for awareness and collaboration in the context of dual-user exhibition viewing scenarios by integrating the results of observational studies. Subsequently, we formulated four design goals to enhance awareness, thereby enabling our system to better support these dual-user collaborative viewing scenarios. In this section, we describe how these goals are achieved within our system.}
\subsubsection{Design Goal 1: Enhance the environmental visibility of local spaces for remote users to improve spatial awareness.}
We developed a basic teleoperation system that allows remote users to control the robot in the local space (i.e., forward and backward, left and right movements) using keyboard WASD inputs. ~\ryrevised{Our observational studies indicated that visual perception is a primary method for viewers to acquire environmental information; thus, any approach that provides a broader field of view aids in achieving this design goal. Previous research on telepresence robots has adopted various methods to enhance environmental visibility for remote users, including the use of wide-angle cameras on the robot side \cite{jia2021tiui}, deploying multiple cameras \cite{jia2021tiui,jones2021belonging}, and transmitting 360° panoramic videos \cite{heshmat2018geocaching}. Drawing from these approaches, we equipped the TeleAware robot with a 120-degree wide-angle camera as its main camera. Additionally, we installed a binocular surveillance camera on top of the robot to provide two new environmental awareness perspectives for remote users. Observational studies have shown that viewers often perceive their environment by scanning up and down and looking around. Therefore, we employed a binocular surveillance camera system consisting of a fixed camera and another camera with an electronic pan-tilt mechanism. Users can remotely control the electronic pan-tilt to rotate the camera 343 degrees horizontally and 120 degrees vertically, enabling up-and-down and side-to-side scanning for visual searching. We aim to assist remote users in surveying their surroundings through multiple, freely controllable camera perspectives, capturing more of the robot's body and the ground around it, and thereby enhancing remote users' ability to cognitively orient themselves and view exhibits.}
\subsubsection{Design Goal 2: Support remote users in perceiving the location and status of
local users to improve social awareness.}
\begin{figure}[ht]
    \centering
    \includegraphics[width=0.8\linewidth]{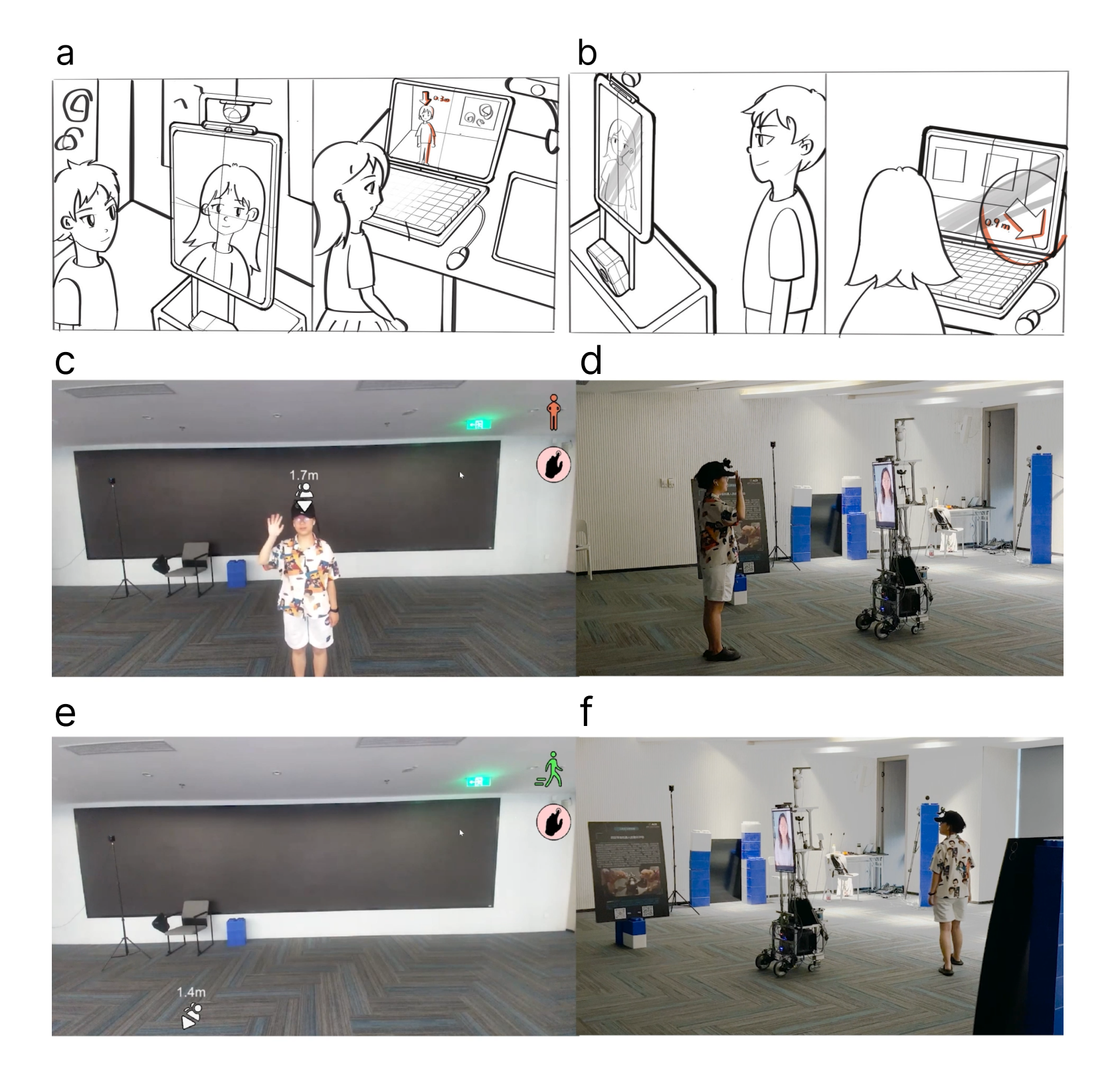}
    \caption{Design concept and prototype effect diagrams for Design Goal 2. a) When the local user appears in the remote user's field of view, the arrow moves to the local user's head, displaying the corresponding relative distance and movement status; b) When the local user is outside the remote user's field of view, the arrow moves to the bottom of the screen, indicating the relative bearing and distance of the local user. c) and d) respectively show the perspectives of the remote and local users when the local user is within the remote user's field of view. e) and f) respectively illustrate the perspectives of the remote and local users when the local user appears in the blind spot of the field of view.}
    \label{fig:designGoal2}
\end{figure}

In the field of telepresence robots, the transmission of 360-degree videos through VR headsets has been utilized to enhance remote users’ perception of local users \cite{heshmat2018geocaching,tang2017collaboration}. However, the transmission of 360-degree videos implies higher demands on network bandwidth and hardware equipment. ~\ryrevised{Some studies have enhanced the perception of partners by framing local users within the field of view of remote users \cite{beraldo2021shared}, yet they lack further design considerations for conditions where partners are in visual blind spots.} Research in other fields indicates that mutual location awareness tools can support implicit coordination and division of labor \cite{dillenbourg1997role}, and mutual understanding of each other's intentions \cite{nova2007collaboration}. ~\ryrevised{Based on this, we developed a location awareness tool centered on the remote users‘ perception to help them understand the position and movement state of their collaborating partner. Trackers were configured on both the robot and the local user to capture and calculate the distance and relative direction between them. This information was then presented on the remote user's screen through directional icons, movement state indicators, and text. We differentiated between blind and non-blind areas, based on whether the local user is within the visual range, to simulate the intuitive perception method of a partner's location in actual collaborative locomotion.} During collaboration, when local users enter the field of view of remote users, directional icons are displayed in real-time above the local users' heads. When the robot's camera cannot capture local users, the directional icons rotate and move to indicate the relative direction of local users concerning the robot. Simultaneously, based on human-shaped icons, remote users can determine their partners' movement state (i.e., stationary, moving). In this way, we aim to enhance the remote user's awareness of the local user. The design concept diagrams and prototype implementation are shown in Fig.~\ref{fig:designGoal2}.

\subsubsection{Design Goal 3: Enhance embodied interaction between local and remote users to improve social awareness.}
\begin{figure}[ht]
    \centering
    \includegraphics[width=0.6\linewidth]{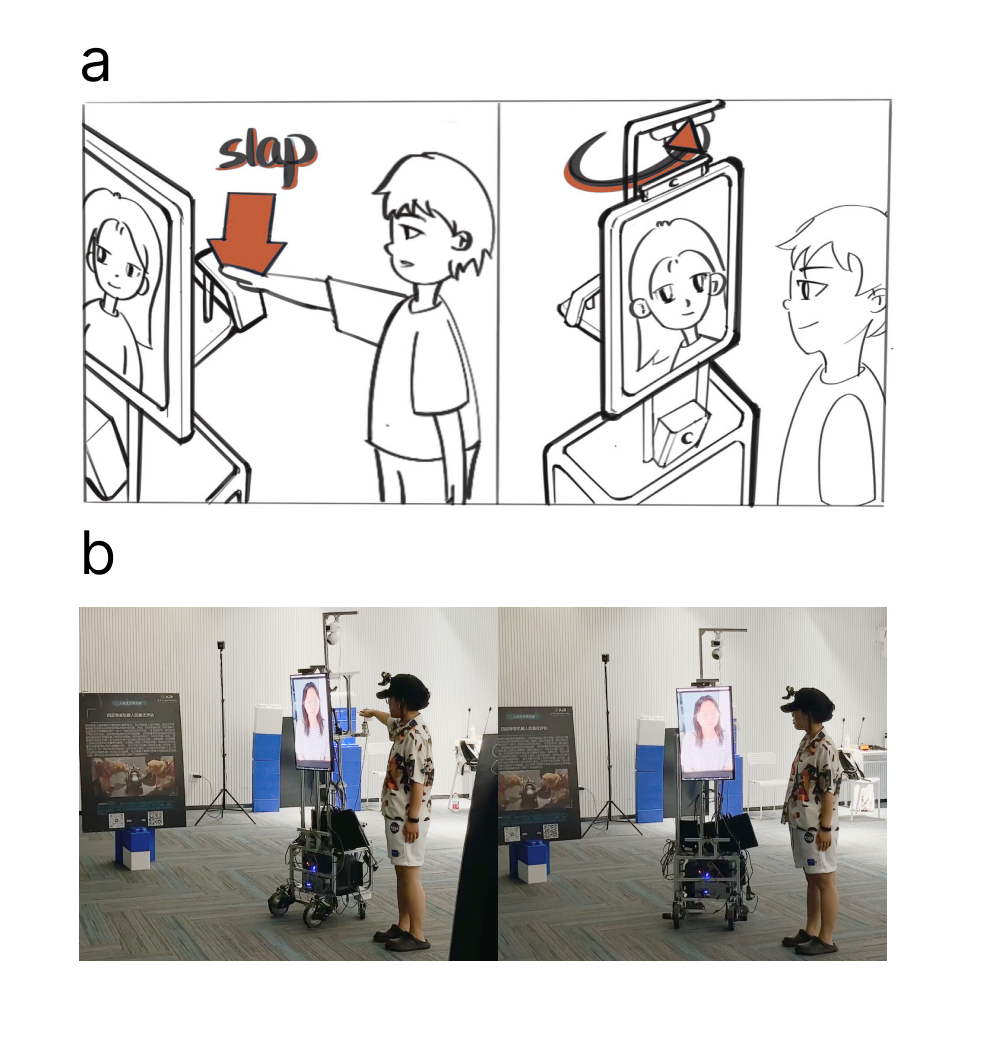}
    \caption{Design concept and prototype for Design Goal 3. a) is a schematic representation of the shoulder-tapping function design; b) shows the actual effect of the prototype feature. When the local user presses the force sensor, the robot automatically turns towards the local user.}
    \label{fig:designGoal3}
\end{figure}

Traditional video call systems often limit themselves to voice and facial expressions, somewhat lacking in body language expression. In the field of telepresence robots, research on embodiment primarily focuses on generating holographic images \cite{orts2016holoportation} or 3D modeling of remote users \cite{jones2021belonging} but lacks representation of intimate interactive actions. Combining results from our observational study, ~\ryrevised{light taps on the shoulder or arm are common interactive behaviors} among collaborators. Considering this, we installed two force sensors on either side of the robot's display screen, acting as the TeleAware robot's shoulders. When a local user presses either the left or right force sensor, the robot automatically rotates a certain angle to the left or right, simulating the natural turning reaction of a person being lightly tapped on the shoulder, thus enhancing tactile interaction. The design concept diagrams and prototype implementation are shown in Fig.~\ref{fig:designGoal3}.

\subsubsection{Design Goal 4: Support joint referencing of environmental information by both local and remote users to improve situational awareness.}
~\ryrevised{In our observational study, we found that gaze following and gesture indication are common behaviors for forming a mutual reference between collaborative partners. }Pointing gestures allow observers and instructors to establish references in communication by positioning to guide visual attention \cite{li2023understanding}. ~\ryrevised{Both parties have equal dominance in expressing their focus and understanding the other's focus. However, remote users, due to the lack of means to express their focus points, often find themselves at a disadvantage when using telepresence robots in collaboration. Therefore, we believe it is necessary to introduce new features to support mutual reference of environmental information, enabling remote users to have an equal role in the collaboration.}

~\ryrevised{For remote users, vision is the key method for understanding on-site users' references. Previous research by Herbort and Kunde revealed systematic errors made by observers in identifying pointing gestures \cite{herbort2016spatial}. They found that although instructors produce gestures using visual-touch lines, observers interpret pointing gestures using the "arm-finger" line. Herbort and Kunde validated that this systematic misunderstanding occurring in human communication is reduced when observers are instructed to extrapolate the touch line vector\cite{herbort2018point}. Inspired by these findings, we utilized a human posture estimation algorithm on the remote user's end to recognize gestures captured by the main camera on the robot's end. We identified pointing gestures based on the angles formed by connected human body key points and presented visual-touch line vectors on the remote user’s screen to improve the interpretative performance of pointing gestures.}

\begin{figure}[ht]
    \centering
    \includegraphics[width=0.8\linewidth]{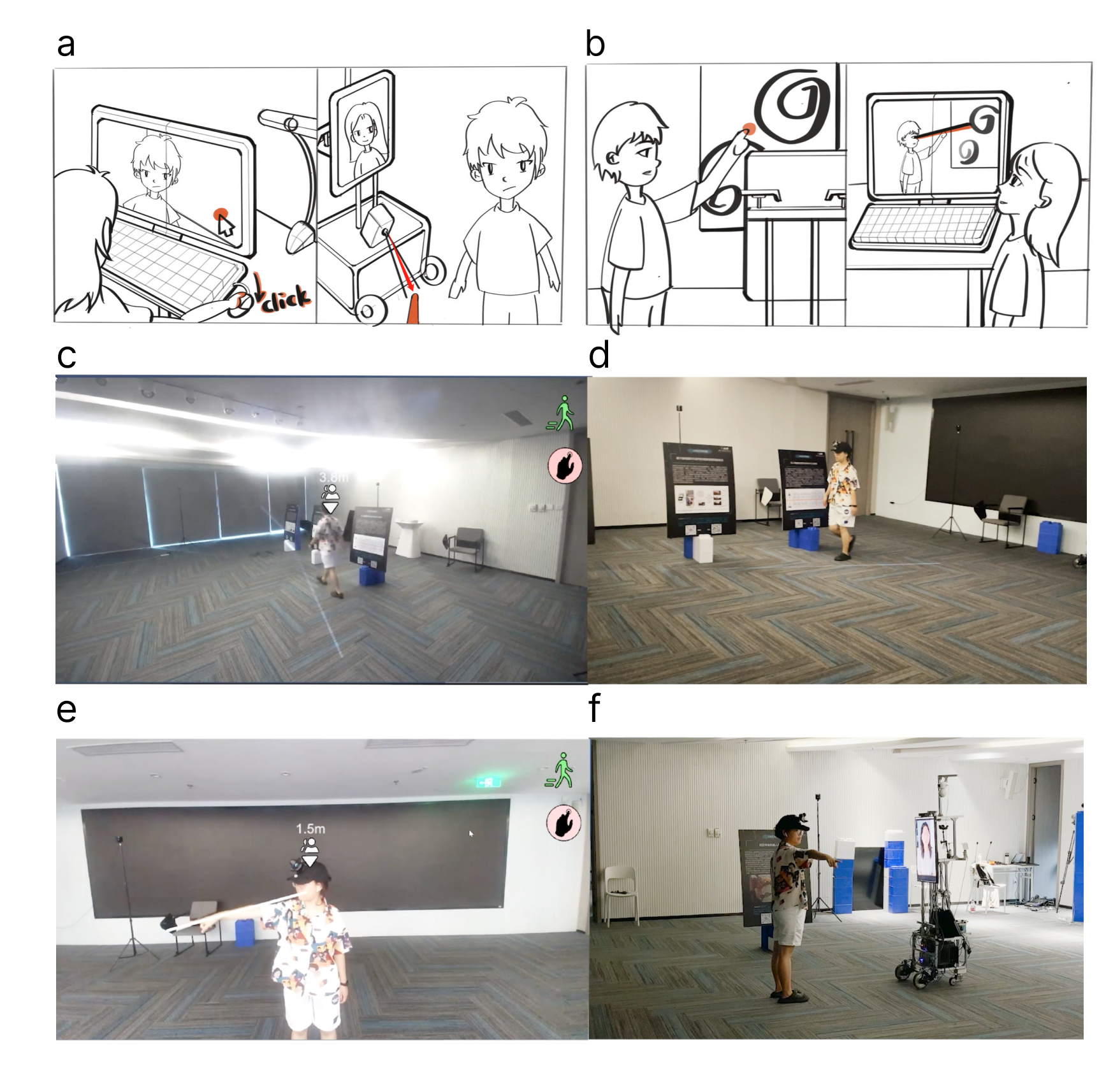}
    \caption{Design concept and prototype effect diagrams for Design Goal 4. a) illustrates the scenario when a remote user clicks on the screen, and the projector on the robot side casts a ray on the ground in the corresponding direction to guide the local user. b) shows the automatic recognition of the local user's indicative posture on the remote user's display, with a guidance line appearing. c) and d) demonstrate the actual effect of the remote user's referential function, where the remote user guides the movement target through a projected beam. e) and f) display the actual effect of the local user's indicative posture recognition, with the guidance line shown on the remote user's end to enhance the prompting effect.}
    \label{fig:designGoal4}
\end{figure}

~\ryrevised{For local users, the concrete representation of remote users' indications can express more comprehensive information. Previous studies have facilitated discussions between remote and local users about content on walls or floors through projector-projected light points \cite{machino2006remote,fischedick2023bridging}and displayed the movement path of robots by projecting arrows \cite{shrestha2018communicating,hetherington2021hey}. Taking inspiration from this method, we installed a projector on the robot. Considering the use of the projector in large spatial areas, it needs to cover references to specific objects and directions. Therefore, we projected ray trajectories onto the floor of the collaborative space for indication. When indicating, remote users use the mouse to click on the direction they want to point on the screen. The system captures the clicked position relative to the display screen and converts it to the coordinates of the projector projection interface. Based on this, remote users can adjust the direction of the ray displayed on the floor of the local space by the projector. This reference design allows both parties to have the dominance in expressing indications, thereby creating an environment of collaborative interdependence. The design concept diagrams and prototype implementation are shown in Fig.~\ref{fig:designGoal4}.}

\subsection{Technical Details}
Based on our design goals and following a series of pilot studies, we have developed TeleAware Robot, a telepresence robot that supports collaborative locomotion. The development of TeleAware Robot is divided into two parts: hardware and software. The hardware part includes the robot body as well as the remote user-side environment construction, and the software part is the two-way communication and data visualization driven by Unity3D.
\subsubsection{Hardware}
The outer frame of the robot body is composed of aluminum profiles, aluminum columns, aluminum hollow tubes, and fiberglass panels. The following is a detailed description of the robot’s composition. (1) two DC brushless geared motors and two brushless motor speed controllers; (2) a ROS main control board STM32F407VET6; (3) two sets of customized polyurethane rubber wheels; (4) a UGREEN 1200W mobile power supply; (5) a HIKVISION DS-UVC-U64 Pro 2K camera; (6) a TP-LINK 600W Binocular Surveillance Camera; (7) two sets of force sensors; (8) a 27-inch REDMI display; (9) a wireless projector; (10) two HTC Vive trackers; (11) a laptop computer to run the robot side software and the electronic control program.

\begin{figure}[ht]
    \centering
    \includegraphics[width=0.8\linewidth]{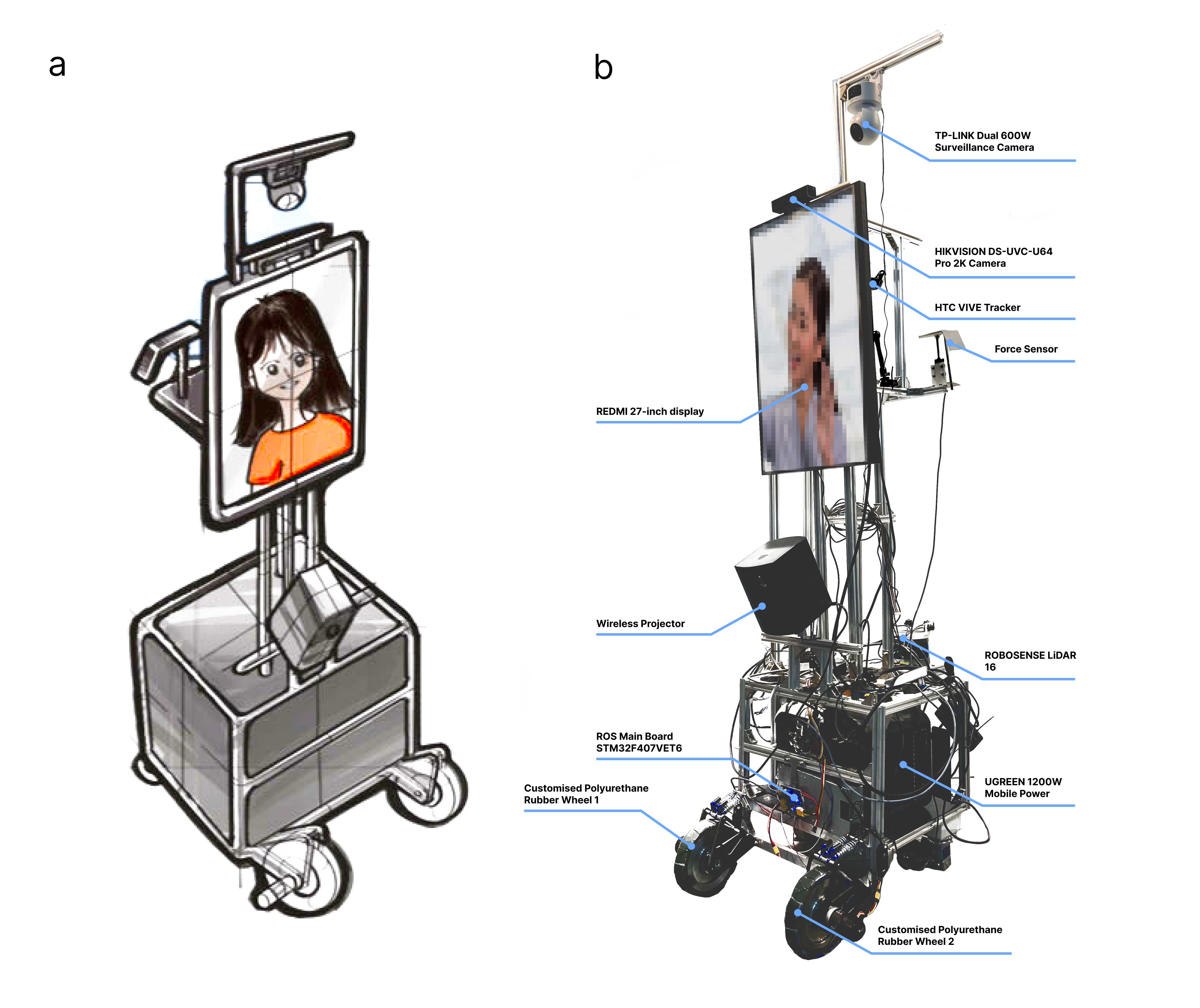}
    \caption{TeleAware Robot. a) Schematic representation of the robot; b) Final implementation of the robot.}
    \label{fig:hardware}
\end{figure}

The remote user side includes the following devices to support remote video and data visualization. (1) a HIKVISION E14a 2K camera; (2) an aluminum desktop stand; (3) an iPad 7; (4) a laptop to run the software.

\begin{figure}[ht]
    \centering
    \includegraphics[width=0.8\linewidth]{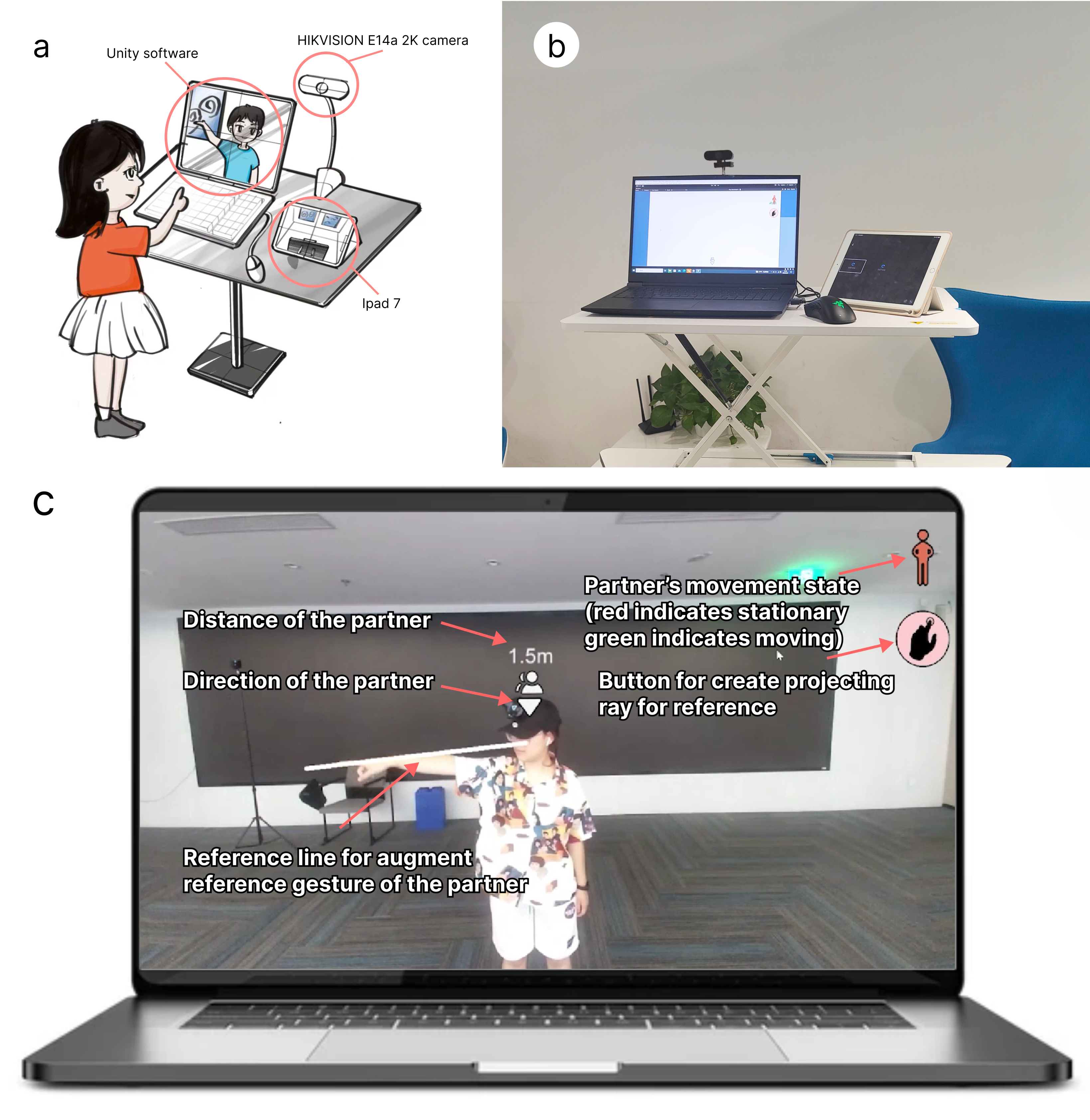}
    \caption{Remote user experimental environment. a) Schematic representation of the physical environment; b) \ryrevised{Final implementation of the physical environment, wherein a laptop is utilized to execute the main program, and an iPad 7 is employed to display the footage captured by the binocular surveillance camera on the robot side; c) Remote user interface running on the laptop.}}
    \label{fig:remoteInterface}
\end{figure}

\subsubsection{Software}
We used the Unity3D engine to develop the software part of the robot because it is less expensive and has great scalability and upgradeability. We used three fully developed and supported software packages for the software part, including the video calling package Agora-RTM, the remote messaging package Agora-RTC, the image recognition algorithm package OpenCV for Unity, and two utilities including the tracker support tool SteamVR, and the projector screen casting software LeboCast.

\textbf{Agora-RTM} video call package is used to provide remote video call functionality. When the robot side and the remote user side join the same video channel, Agora-RTC automatically captures and transmits all user camera frames on the same channel, enabling low-latency remote video calls. In our application, the user automatically joins the same video channel at the start of the program to enable remote video communication.

\textbf{Agora-RTC} Remote Messaging Package is a software package for transferring messages between clients, where all clients joining the same messaging channel are able to receive messages from other users on that channel with low latency. We use this package to complete the message transfer between the remote user and the robot side, including the remote keyboard input control, the pointing position of the remote user clicking on the screen, and the location data of the two HTC Vive trackers in the field.

\textbf{OpenCV for Unity} Image Recognition Algorithm Package supports the OpenCV open-source computer vision library, which enables rapid implementation of computer vision applications under the Unity engine, including target detection, image processing, feature extraction, and so on. In our application, we use YOLOX Object Detection and MediaPipe Pose Estimation to perform human body recognition and human body key point detection for the robot-side camera and display the recognition results on the remote user-side screen in real time for instant and dynamic remote discrimination.

\section{EXPERIMENTAL STUDY}
In our study, two participants formed a pair, with one participant's experimental environment as a local user and the other as a remote user, with each participant's environment remaining fixed throughout the experiment. The two participants played different task roles in each round of the experiment. A 2 $\times$ 2 mixed experimental design was employed, with two main independent variables: the robotic system [TeleAware Robot, Standard Telepresence Robot] and the task role [Leader, Follower]. Below, we briefly describe the experimental tasks and the definitions of factors.

\paragraph{Robotic System.}
The study established two conditions based on the robotic system to assess collaboration quality and effectiveness among different robotic systems during collaborative locomotion tasks.

(1) \textit{TeleAware Robot} - An enhanced version of the telepresence robotic system.

(2) \textit{Standard Telepresence Robot} - The basic version of a telepresence robot. Remote users could control the robot to move in the local space using a keyboard and could see the local space through a 120-degree wide-angle camera. However, this version did not include the added binocular surveillance cameras or other enhanced technologies.

\paragraph{Task Roles.}
Two task roles were established in the experiment.

(3) \textit{Leader} - Move according to a route map.

(4) \textit{Follower} - Follow the leader during the movement and try to remember as much environmental information as possible.
\subsection{Motivation}
The design of the experimental tasks was based on the ~\ryrevised{spacious exhibition environment observed in the observational study.} The experimental conditions and sequence were fully counterbalanced. The tasks included two roles: a leader and a follower. The leader was required to move and explain according to a given route map, while the follower had to move following the leader and remember as much information from the display boards as possible. ~\ryrevised{One of the reasons for this design is that it aligns with the framework derived from our observation study, where the core of dual-user collaborative exhibition viewing is the "User-Partner-Environment" interaction. While the observation study involved exhibition scenes with varying visitor flow, the degree of crowding did not alter the core framework or awareness requirements. Furthermore, among the data of the 16 participant pairs analyzed quantitatively in the supplementary study, we noted only one instance of verbal communication between a pair and individuals outside their group. Hence, the experimental task involves only two viewers, excluding other social roles.} Another reason is the previous research has shown that collaborative locomotion activities such as exhibition viewing involve various modes of movement, among which "leading" and "following" are common methods of distributing movement authority \cite{dim2014automatic}. Additionally, some studies \cite{yang2018shopping} mention that telepresence robots, due to their lower mobility and narrower field of view, struggle to act as leaders in collaborative locomotion scenarios. Based on this, the design of the experimental tasks aimed to simulate the distribution of movement authority in ~\ryrevised{collaborative exhibition viewing} and explore whether the enhanced TeleAware robot system could effectively fulfill the role of a leader. This motivated our design of the experiment task.
\subsection{Participants}
A total of 24 participants, aged between 21 and 48 years (M = 26.25, SD = 7.4, 10 males, 14 females), took part in our study. Participants were recruited through electronic posters distributed to social groups. The poster included an overview (i.e., a collaborative telepresence robot task with an exhibition theme), time, location of the experiment, and information about participant compensation. A QR code linked to an online questionnaire was attached to the poster. Interested individuals were required to fill out the questionnaire. The first part of the questionnaire contained demographic information about the participants, while the second part included information relevant to the experiment, such as participants' experience with remote conferencing systems, how often they use remote conferencing systems for collaboration, their knowledge of robot control, attitudes towards smart technology, and their needs for existing remote conferencing systems. All participants who took part in the experiment signed an informed consent form. At the end of the experiment, participants were compensated for their participation.
\subsection{Task}
Based on the results of our observational study, we designed the overall environment of the experiment to mimic an exhibition hall. A total of eight display boards were prepared, with six placed in the experimental space for each round of the experiment, of which four contained text and images, and the other two were blank. The experimental tasks included two roles: Leader and Follower. In each round of the experiment, the two participants in the group were asked to play the roles of Leader and Follower, respectively. To allow both participants to experience the two task roles, each group underwent two (task roles) $\times$ two (robotic systems) = four rounds of experiments. Additionally, we designed four different panel layouts and guided tour routes for each round of the experiment to minimize learning effects. During the experiment, the Leader had to move according to the guided tour map provided in each round, leading the Follower to view the display boards. While following the Leader, the Follower had to try to remember the main content of each display board and its location in the space. At the end of each round of the experiment, the Follower was required to fill out the location and content of each display board on a blank scene map questionnaire. 
\subsection{Experimental Environment}
~\ryrevised{Considering the spatial requirements of collaborative locomotion,} we designated an 8 $\times$ 8-meter indoor experimental area in the center of a 9 $\times$ 10-meter lecture hall, providing participants with a sufficiently large area for movement. HTC VIVE base stations were installed at the midpoint of the boundaries of the experimental area (8 $\times$ 8 meters), at a height of two meters above the ground, to cover the entire area. Subsequently, two trackers were mounted on the hat worn by local users and on the robot, which could receive infrared light signals emitted by the base stations and provide feedback. Using HTC Vive Trackers, we were able to obtain real-time positional coordinates of both local users and the robot. A TPLINK IPC55A panoramic camera was installed in the northwest corner of the experimental area, two meters above the ground, to record the entire experimental process. Additionally, a GoPro camera was set up on the remote user's end to document their operations during the experiment. Throughout the experiment, an observer was stationed outside the experimental area for observation and recording purposes.

\begin{figure}[ht]
    \centering
    \includegraphics[width=0.8\linewidth]{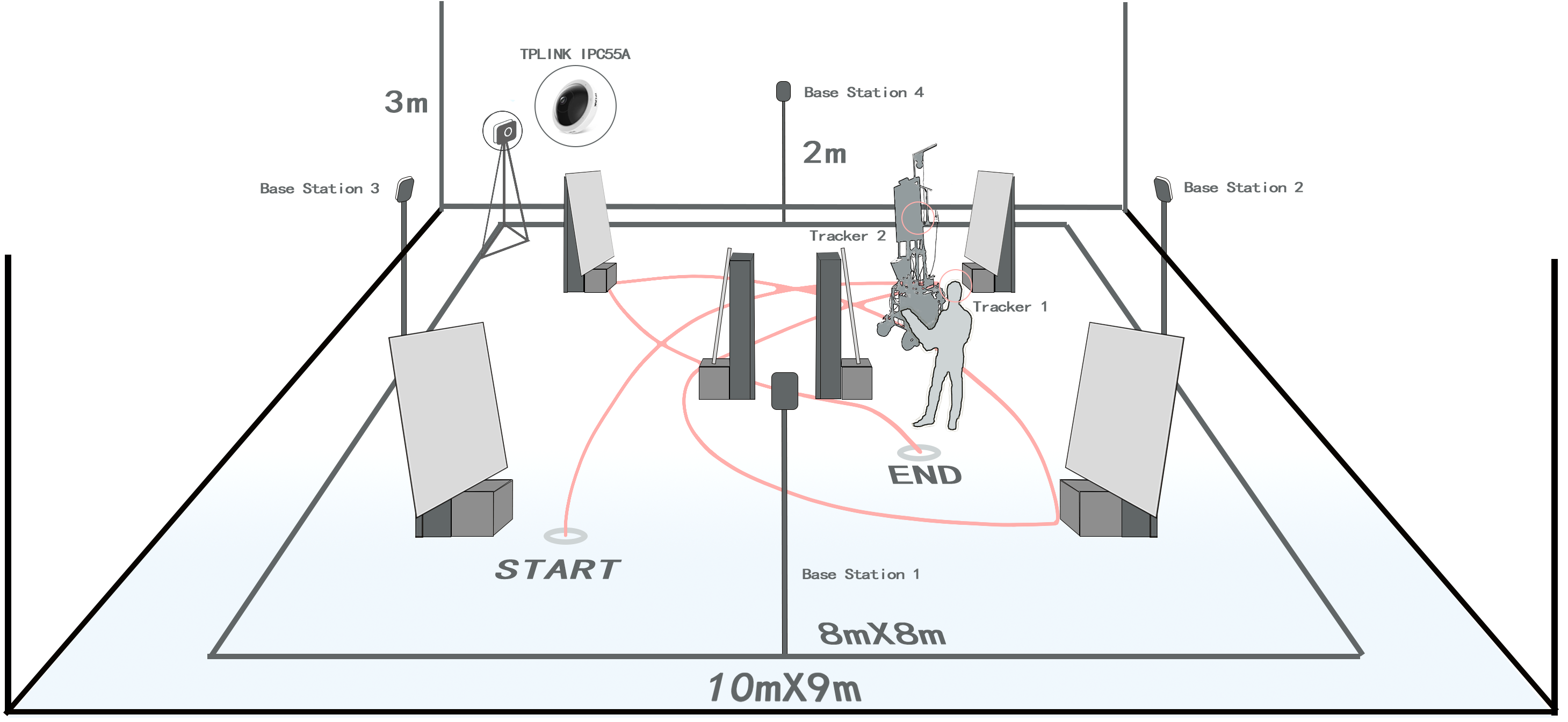}
    \caption{Schematic diagram of the experimental environment in the lecture theatre. Leader and follower paths are drawn as the demonstrations.}
    \label{fig:e-environment}
\end{figure}

\subsection{Procedure}
Participants were randomly grouped into pairs. After grouping, the experimenter designated their experimental environment settings (i.e., remote or local users), and each subject's experimental environment remained fixed throughout the process. All participants were required to complete both training and the formal experiment. During the training period, we introduced the operation of the robot and the experimental rules and allowed participants to practice operating the robot's functions. The formal experiment lasted approximately 60 minutes. To avoid order effects, task sequences were fully counterbalanced among participants. The sequence of the robot system phase was counterbalanced by switching, followed by counterbalancing the task role sequence in each robot system scenario using a Latin square among the pairs. To minimize random factors affecting the experimental results, each pair of subjects performed in each scenario of robotic systems (TeleAware, Standard) $\times$ task roles (Leader, Follower). Before each round of the experiment, we informed the subjects of their task roles for that round and provided the leader with a corresponding navigation route map. The system recorded the real-time positions of the local user and robot, as well as the duration of the experiment. At the end of each round, the follower first completed a blank scene map questionnaire, after which both subjects filled out the Social Presence Scale, NASA-TLX Scale, and IOS Scale. To gather information about the subjects' cognitive differences in different experimental settings during the interviews, at the end of four rounds of experiments, the two subjects switched experimental settings (i.e., from remote to local user and vice versa), with the new remote user acting as the leader for a round of experience, which was not included in data collection. Structured interviews were conducted after all the experiments were completed. The entire process was video-recorded.
\subsection{Measures}
This study employed both objective and subjective performance measures for task performance, task load, social proximity, social presence, and system experience.

(1) \textbf{Log Data.} To assess whether the enhanced prototype impacted the performance of the collaborative tasks and user mobility, we recorded the time taken by the groups to complete the tasks and their movement trajectories within the venue.

(2) \textbf{Memory Questionnaire Data.} Blank scene map questionnaires were prepared to evaluate the followers' memory and perception abilities of the environment.

(3) \textbf{Video Data.} To analyze if there were any specific behavioral patterns among collaborators during the experiment and the behavioral differences brought by different task roles, the entire experimental process was recorded by video cameras, including the GoPro camera in the remote user's experimental space and the TPLINK IPC55A panoramic camera placed at the local space.

(4) \textbf{Measurement Scales.} Different scales were prepared to assess the social presence, social connectedness, and cognitive load. After each of the four rounds of experiments conducted by each group of subjects, both participants were required to fill out the Social Presence Scale, NASA Task Load Index (TLX), and the Inclusion of Other in Self (IOS) Scale. The Social Presence Scale \cite{biocca2001networked} consists of three parts: co-presence, psychological involvement, and behavioral engagement, to assess social presence. Hart and Staveland's NASA-TLX method \cite{hart2006nasa} assesses workload using five 7-point scales. Additionally, the IOS Scale \cite{aron1992inclusion} was used to access social closeness, allowing subjects to select a pair of overlapping circles from seven different choices to describe their relationship with their partner.

(5) \textbf{Interviews.} To further obtain insights into the subjects' experiences and perspectives on using the TeleAware prototype, interviews were conducted with all participants regarding their system and task role experiences. They were interviewed after four rounds of experiments and one round of experience. In the system experience section, we inquired about the users' preferences for the system, as well as their feelings and suggestions for the four design goals of the TeleAware robot. Regarding the task role experience, we mainly asked about the differences in experiences as a follower and a leader and the perceived differences in conducting experiments in different experimental environments. The content of the interviews was fully recorded and transcribed for further data analysis.
\subsection{Findings}
\subsubsection{Task Performance and Task Load}
We compared the overall performance of subjects under the TeleAware robot condition and the standard telepresence robot condition, including task completion time and accuracy of followers in terms of the memory display board. We found no significant difference in overall performance between the two system conditions.

Regarding task load, we used Hart and Staveland's NASA-TLX scale to assess the overall workload and work experience of users under different robot conditions. The results showed that the TeleAware group perceived significantly lower cognitive demands (p=0.046, t=2.017) and less frustration and negative emotions (p=0.040, t=2.076) compared to the standard condition. These findings suggest that our TeleAware robot effectively reduced the perceived task load and improved user experience.
\subsubsection{Navigation Patterns and Social Proximity}
\begin{figure}[ht]
    \centering
    \includegraphics[width=0.9\linewidth]{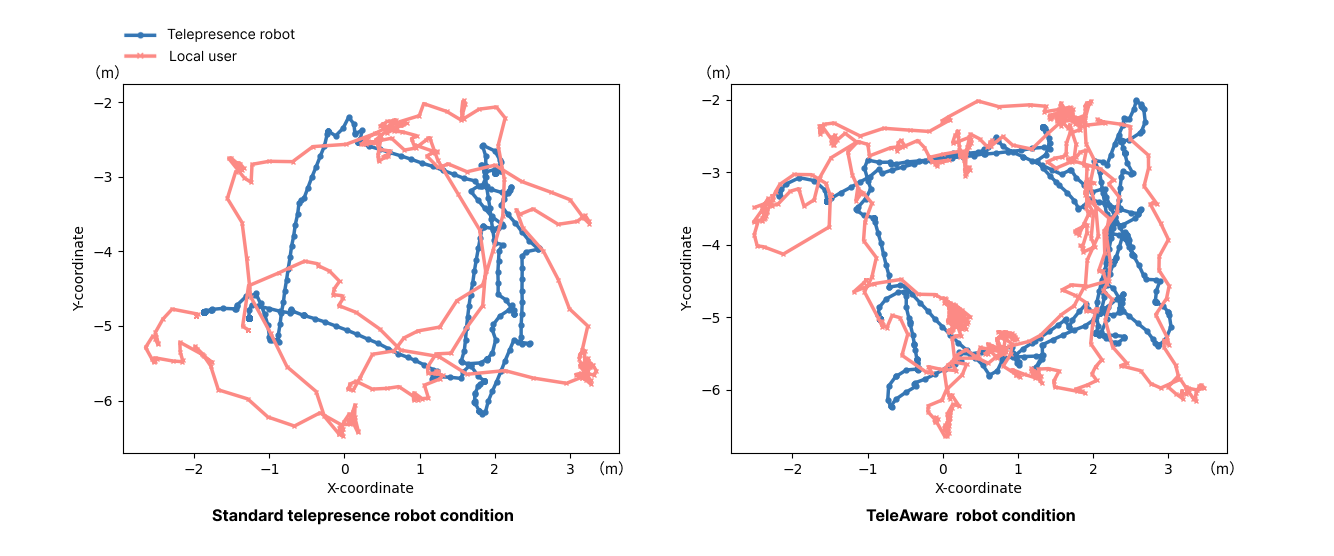}
    \caption{Examples of the trajectories of the telepresence robot and the local user under two conditions. Compared to the standard condition, trajectories under the TeleAware condition exhibit higher overlap and closer average distance.}
    \label{fig:trajectory}
\end{figure}

In terms of navigation patterns and social proximity, after conducting normality tests, we used the t-test to calculate data from the IOS Scale. Both remote and local users under the TeleAware condition reported significantly higher scores (remote user: p=0.006, t=2.879; local user: p=0.037, t=2.145), indicating a stronger sense of social and emotional connection when using the enhanced prototype. Additionally, according to the previous research\cite{yang2018shopping}, the physical distance between collaborators can be an indicator of social proximity. Due to data transmission issues, tracker data for other groups was not fully recorded. Hence, we only calculated the average distance between the local user (as a leader) and the robot within the same group of subjects (N=5). The average distance in the TeleAware group was 1.21 meters, compared to 1.48 meters in the standard condition group. The significantly smaller average distance in the TeleAware group (p=0.046, t=-2.849) indicates closer collaborative locomotion between the local and remote users under the TeleAware condition, as exemplified in Fig.~\ref{fig:trajectory}.

\subsubsection{Mutual Awareness and Social Presence}
\begin{figure}[ht]
    \centering
    \includegraphics[width=0.8\linewidth]{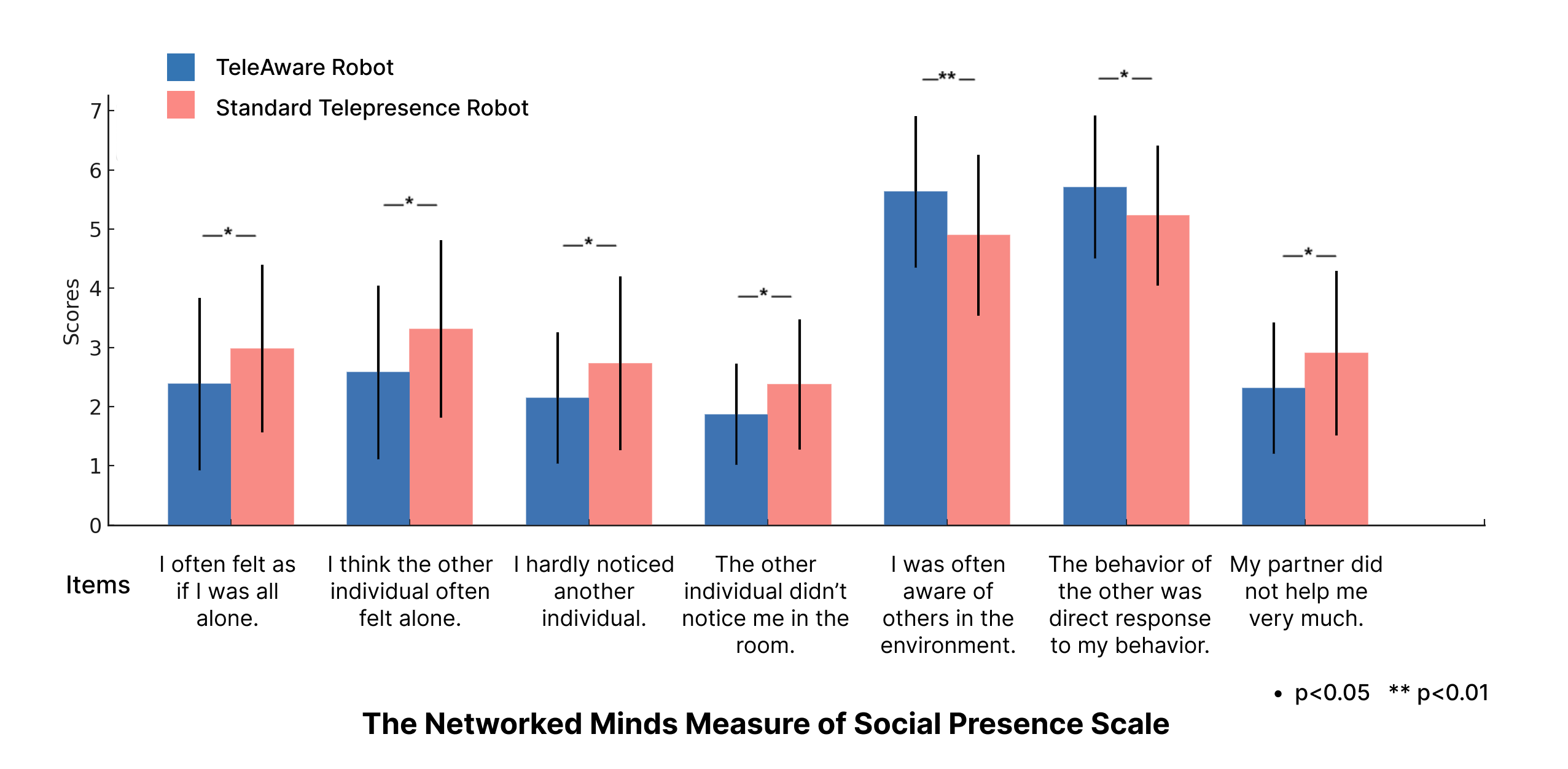}
    \caption{This figure presents the scores of items in the social presence scale that showed significant differences. The data indicate that users under the TeleAware condition have a higher sense of mutual awareness and social presence compared to the standard condition.}
    \label{fig:scaleChart}
\end{figure}

Data from the Social Presence scale was analyzed and passed the normality test. Based on this, we conducted t-tests comparing the scale results under the TeleAware robot and standard conditions. The Social Presence scale sub-scale measuring mutual awareness showed significant results in all sub-items for users of the TeleAware robot. Notably, the mutual awareness scale included several questions assessing different aspects of awareness. In terms of loneliness, TeleAware users reported lower scores in feeling lonely and isolated compared to standard users (p=0.037, t=2.112) and were more inclined to feel that their partner was not alone (p=0.014, t=2.505). In terms of social attention, TeleAware users reported higher scores, indicating closer attention to the partner (p=0.026, t=2.259). Also, compared to the standard group, participants in the TeleAware group reported significantly higher scores on feeling the presence of the partner (p=0.006, t=-2.816). Furthermore, compared to standard conditions, users of the TeleAware robot perceived their companion's responses as more direct (p=0.043, t=-2.05) and the help provided by their companion as better (p=0.018, t=2.415). The above results are shown in Fig.~\ref{fig:scaleChart}.

Overall, the results indicate that our designed TeleAware robot effectively enhanced mutual awareness and social presence among collaborators, fostering the occurrence of collaborative behaviors.

\subsubsection{Leader and Follower}
As mentioned in our experimental design, the task design simulated the distribution of movement authority in collaborative locomotion. We aimed to explore whether a telepresence robot could act as a leader and what impact this role would have. Previous research has discussed the impact of this type of authority distribution to a lesser extent, so we also analyzed the task roles (leader and follower). Overall, local users are indeed well-suited to act as leaders, but remote users using the telepresence robot can also collaborate effectively through new behavioral methods.~\ryrevised{Furthermore, the use of the TeleAware system enables remote users to feel less isolated when acting as a leader.}

By conducting t-tests on questionnaire data, we found that leaders tended to make more effort to attract their partner's attention (p=0.008, t=-2.705) and felt that their behavior had a more significant impact on the follower’s actions (p=0.001, t=-3.603). Meanwhile, followers reported a higher degree of dependence on the leader (p<0.001, t=4.418). ~\ryrevised{Additionally, remote users as leaders completed tasks in an average time of 276.2 seconds, while local users as leaders completed tasks in an average time of 327.5 seconds. This indicates that remote users can act as leaders and complete tasks more quickly in that role. According to behavioral observations, in the absence of higher mobility, remote users actively used verbal communication and projection references to inform local users about the next display board, while local users with higher mobility reached the display board ahead of time and started reading its contents. This became a strategy among collaborators, allowing the robot leader to play a more substantial referential role. Finally, under the TeleAware condition, compared to standard conditions, remote users as leaders felt less isolated. Data from the Social Presence Scale indicated that under the TeleAware condition, remote users as leaders felt the presence of the local users around them more (p=0.002, t=3.553). Additionally, P19 in the remote environment commented: "Without this feature, when my companion wasn’t in my line of sight, I felt like I was watching the exhibition alone, a bit uncertain about what to do next."}

Overall, the inherent perceptual and mobility capabilities of local users naturally give them an advantage when they assume the role of a leader. ~\ryrevised{However, telepresence robots are still capable of adopting the role of a leader and can participate in collaborative locomotion through innovative behavioral methods. Furthermore, the enhanced TeleAware system can mitigate the sense of isolation experienced by remote users when acting as leaders, facilitating the completion of collaborative locomotion in a more coordinated manner.}

\subsubsection{Preferences and Reasons}
After the experiment, subjects indicated their preference between the two systems. ~\ryrevised{We conducted an induction and organization of the interview content. Initially, we transcribed the interview content into text to ensure the completeness and accuracy of the data. The data underwent preprocessing to retain effective content for analysis. We categorized the interview content, summarizing it according to four design goals. Two members participated in the coding of the interview data. These members, through careful reading of the interview text, identified statements or paragraphs expressing the same or similar meanings and conducted preliminary labeling or coding. Based on this, we analyzed the results of the open coding, organizing key meanings to discern users' thoughts and opinions regarding the implementation involved under each design goal.} 

In terms of preference, the TeleAware robot system outperformed the standard telepresence robot system. 87.5\% of participants preferred the TeleAware robot system, 8.3\% saw no significant difference between the two systems, and 4.2\% felt that the standard one was sufficient for achieving the objectives. The following user feedback is organized according to the design goals presented in Section~\ref{Sec:Findings}.

\paragraph{\textbf{Design Goal 1: Enhance the environmental visibility of local spaces for remote users to improve spatial awareness.}}
Most remote users noted that the rotatable binocular camera mounted on top of the robot had a positive impact on the robot's mobility. Compared to the wide-angle camera, distortion-free surveillance camera footage better assisted in judging distances, avoiding obstacles, and providing a clearer view. For instance, P3 mentioned:

\begin{quote}
    \textit{"I often looked at the auxiliary camera footage while moving because it could capture the robot's body, and I could understand how wide the robot actually was."}
\end{quote}

However, 8.3\% of participants mentioned that shifting their gaze from the main display to the tablet displaying the binocular camera footage posed a challenge for them. Additionally, since operating the remote rotatable binocular camera required actions on the tablet, this often led users to stop moving to specifically adjust their viewing angles. 

Overall, remote users found that the footage provided by the rotatable binocular camera was more effective for them in clearly perceiving the environment and to some extent, could simulate looking around behaviors. Yet, they still needed to be mindful of the difficulty in switching their gaze between multiple displays.


\paragraph{\textbf{Design Goal 2: Support remote users in perceiving the location and status of
local users to improve social awareness.}}
 Regarding understanding the partner's location and movement state, 83.3\% of participants stated that displaying the partner's location and relative distance was crucial when the partner was not visible. Especially when the remote user acted as a leader, mutual location-awareness tools eliminated the unknowns about the local user, promoting behavior formation closer to co-located collaboration. P4 even decided whether to start explaining the contents on the display board based on the icon indicating whether the partner had stopped moving. 

Additionally, users reported that enhanced display of partner location helped reduce the number of times they had to maneuver the robot to turn around and look for the partner, easing the remote user's operational burden (P2, P4, P10, P18). For local users, when the remote user could understand their location and movement state, they felt that they would communicate their location verbally less and felt safer. For example, P2 said:

\begin{quote}
    \textit{"Without this feature, if I stood on one side of my partner, I would have to explain to him which side I was on. But with this system, I don't have to say it because he can know where I am. And I can be closer to the robot because the remote user knows where I am, so he won't hit me when operating the robot."}
\end{quote}

Overall, the mutual location-awareness tools under Design Goal 2 effectively supported collaborative locomotion in the user-partner relationship and also showed improvements in social connection and remote operation.

\paragraph{\textbf{Design Goal 3: Enhance embodied interaction between local and remote users to improve social awareness.}}
When discussing the shoulder-tapping feature, subjects considered shoulder-tapping a more interesting feature, improving the expression of non-verbal hints and closer to natural human reactions (P7, P9, P15, P21). They also thought that shoulder tapping could enhance social attention between local and remote users. For example, P21 said:

\begin{quote}
    \textit{"I feel especially when I'm behind the robot, it's just like I normally tap someone on the shoulder, just to let them know I'm behind them."}
\end{quote}

On the other hand, some remote users reported that using the shoulder-tapping feature while performing tasks was challenging for them, as sometimes they couldn't feel the local user tapping on the shoulder because the tool didn't provide them with enough feedback. 

Overall, user feedback proved that the shoulder-tapping function has room for further deepening, but the implementation of the prototype can already provide an effective way for the formation of social attention, and even bring new fun to social interaction.

\paragraph{\textbf{Design Goal 4: Support joint referencing of environmental information by both local and remote users to improve situational awareness. }}
In terms of enabling both parties to achieve shared attention and reference, most subjects felt that the projector projecting directions on the floor accelerated communication, was referential, and conveyed information. For instance, P7 mentioned:

\begin{quote}
    \textit{"With guidance on the ground, I can more clearly know which direction my companion is pointing to. This can eliminate a lot of unnecessary communication without using words."}
\end{quote}

On the other hand, a minority of participants thought these projections were too light in color, making them hard to see. Additionally, participants reported that local users' referential posture recognition caught their attention and enhanced their understanding of the reference direction. However, due to algorithm recognition issues, the display time of the referential guide line on the screen might be too short (P8). 

Overall, while the design itself has room for further iteration, the use of projector ray projection and posture recognition of local users effectively enabled both parties to form a shared reference. \ryrevised{This aspect of the design can enhance the understanding of the space. Additionally, it can improve the perceptual efficiency of forming a common focus, thereby facilitating the completion of leading-following tasks. }

\section{Discussion}
\subsection{Design to Support Collaborative Locomotion Activities}


Dyadic collaborative exhibition viewing represents a typical task within collaborative locomotion activities. Our observational study, conducted within the "User-Partner-Environment" framework, targeted this specific scenario and employed video coding to analyze viewer behaviors. We discovered that completing collaborative locomotion tasks necessitates addressing a multitude of awareness needs encapsulated by this framework. In the user-environment relationship, there exists a substantial demand for a clear perception of the environment, with the visual channel emerging as the predominant method for acquiring environmental information, thereby improving spatial awareness. In the realm of user-partner relationships, achieving social synchronization necessitates continuous information gathering about the partner and subsequent action. Visual perception and physical contact stand out as effective means to glean information about and ascertain the status of partners, with female-female pairings displaying a higher propensity for physical contact. These behaviors help users maintain and enhance social awareness, promoting the formation of leadership. When considering the user-(partner-environment) nexus, we observed active engagement in forming environmental references through gestures and gaze following. This reciprocal exchange of referential information between viewing parties not only augments situational awareness but also smoothens the transition and exchange of leadership roles between individuals.

~\ryrevised{Based on the results of our observational study, we established design goals and undertook design practice. }Through the analysis of qualitative and quantitative data, the system design of the TeleAware robot, although not significantly improving task completion time and answer accuracy, effectively reduced the task load in collaborative 
 locomotion scenarios and improved the user experience. At the same time, the enhanced system significantly boosted mutual awareness and co-presence among collaborators and promoted closer social connections. 
The implementation strategies under this framework effectively supported the three parts of the relationship: user-environment, user-partner, and user-(partner-environment), enhancing users' collaborative awareness and achieving the four set design goals. We believe that TeleAware effectively supports dyadic collaborative locomotion scenarios.

From the subjective experiences of participants, we learned about user preferences for the prototype and their perceptions of the design solutions. ~\ryrevised{Multi-camera perspectives with rotatable views increased remote users' visibility of the local environment, aiding comprehensive environmental perception.} Mutual location-awareness tools reduced the remote user's sense of loneliness and were effective in collaborative locomotion. Enhanced embodied communication provided collaborators with social interactivity and fun, although participants felt that the feedback received by the remote user needed further enhancement. Support for shared reference improved communication accuracy and can continue to be enhanced in terms of projection display.

Overall, the study showed the effectiveness of our approach and framework. ~\ryrevised{We believe that TeleAware effectively supports awareness needs within the dyadic collaborative exhibition viewing scenario and provides design insights for designing telepresence robots in collaborative locomotion activities.}

\subsection{Awareness and Leadership in Dyadic Collaborative Locomotion Activities}
~\ryrevised{Through observational studies, we discovered that both parties in dyadic collaborative locomotion activities possess leadership, and leadership can dynamically transition between them. Completing collaborative locomotion activities requires various types of awareness, and the formation and handover of leadership occur under the influence of these awareness types. In terms of spatial awareness, understanding of the space is a prerequisite for locomotion. Users obtain sufficient information about the space through visual channels, thereby planning their collaborative path. Regarding social awareness, the collaboration in locomotion is based on the understanding between the user and their partner about each other's social context. Once the location and state of the companion are understood, viewers can promptly adjust their behavior, forming social synchrony with their companion. Among these, visual perception of the partner is the most direct way, knowing where the partner is directly influences how users move. Physical contact is an effective action to evoke social attention. By touching the partner's shoulder or arm, one party often follows the other to view the relevant information, thereby forming different leading roles. In terms of situational awareness, knowing what the partner is focusing on can become the driving force for the formation of a collaborative movement. Through gesture pointing and gaze following, the roles of follower and leader quickly transition. Shared reference supports the dynamic transition of leadership.}

~\ryrevised{Previous studies involving telepresence robots have seldom discussed the switch of leadership. A few studies explicitly mentioned that remote users often find themselves in the following position when using telepresence robots, only taking on leadership under compelling reasons \cite{yang2018shopping}. Our experimental data revealed that remote users are capable of assuming a leadership role, with telepresence robots as leaders completing tasks in less time than local users acting as leaders. The TeleAware system, equipped with awareness-augmenting functionalities, can mitigate the sense of isolation experienced by remote users when assuming leaders and provide them with an expanded toolkit for comprehending and conveying their referential information. This supports participants in more effectively accomplishing collaborative locomotion tasks.}

~\ryrevised{This insight inspires us that future designs of telepresence robots for collaborative locomotion should thoroughly incorporate the awareness needs within the "User-Partner-Environment" framework. By enhancing the essential awareness, both remote and onsite users can be endowed with equal leadership capabilities, thereby facilitating collaborative locomotion more effortlessly. This represents a valuable design guiding principle worth considering.}

\subsection{The Impact of Location Awareness on Territoriality}
Territoriality has been widely discussed in the literature on collaborative activities. It naturally forms, helping people mediate their social interactions by linking space with individuals in computer-supported cooperative work \cite{ruddle2015performance,scott2004territoriality,sigitov2019task}.

Under the standard robot condition, we observed that local users typically maintained a greater distance from the robot. Both parties preserved their larger physical territories to prevent safety issues like collisions, similar to findings in previous studies in remote shopping scenarios \cite{yang2018shopping}. One reason for this behavior was the prototype robot's wheel system and network latency limitations, which still rendered the robot's movements somewhat insensitive. ~\ryrevised{Particularly in dual-user collaborative exhibition viewing scenarios, which involve high mobility,} this lack of mobility made it necessary for local users to maintain a certain distance to observe the robot's movement trends. Another reason was that under standard telepresence robot conditions, the robot's ability to express non-verbal hints was significantly weakened when the local user could only see the robot's back or side. This was particularly evident in situations characterized by frequent turns and movements. For example, unable to determine if the robot might suddenly turn, users chose to maintain a greater distance from the robot to ensure their safety. Under this condition, local users had lower trust in the robot's movements, leading to lower territorial overlap.

Compared to the standard condition, we observed that under the TeleAware condition, local users maintained a closer mode of locomotion with the robot. They shared space, achieving a higher degree of territorial overlap. We supported this finding with various indicators, including a significantly higher degree of trajectory overlap between local users and the robot, a significantly closer average distance (Fig.10), and a stronger and closer social connection. Our results suggest that enhanced remote user awareness of the local user's positional state brought about this change. On one hand, location awareness influences the remote user's way of moving. The remote user could more accurately understand the companion's position, especially when the companion was out of sight. The elimination of the unknown state allowed the remote user to reduce the psychological pressure of causing collisions and more freely and accurately change position and direction. On the other hand, location awareness enhanced the local user's trust in the robot's movement. When local users knew that even if they were not in the robot's field of vision, the remote user could still understand their position throughout, they developed higher trust in the robot's movement. This trust made shared territory possible, promoting physical and psychological intimacy, and thus a higher degree of territorial overlap.

\begin{figure}[ht]
    \centering
    \includegraphics[width=0.8\linewidth]{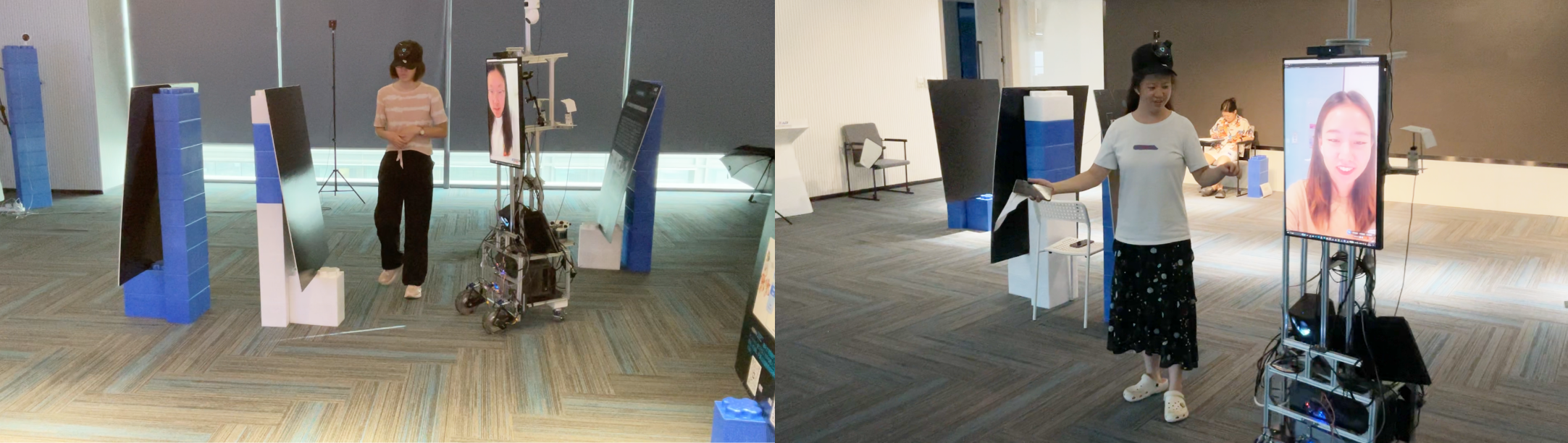}
    \caption{Local users interact with remote users.}
    \label{fig:localuser}
\end{figure}

\subsection{Design Implications}
Reflecting on our design goals and implementation strategies based on the experimental results, we identified several design opportunities worth extending in the future, as well as considerations in technology and design.

\subsubsection{Improving Visibility for Remote Users.}
Improving visibility includes not only the visibility of physical objects in the environment for remote users but also the visibility of their volume. In real collaborative locomotion, users need to move freely in complex environments and view real physical scenes and objects, including texts, images, etc. In our tasks, when remote users played the role of followers, they often struggled to discern content on display boards. Limited by environmental lighting and camera clarity, remote users had to constantly adjust the robot's angle and distance to clearly see this information. This prevents collaborators from providing timely remote driving operations based on accurate information~\cite{chu2023work}. Adding cues about the robot's volume can help users move to a closer position to view physical objects. ~\ryrevised{Meanwhile, cameras with rotational capabilities allow users to adjust their field of view, acquiring more comprehensive information about their surroundings.} Existing image and video processing technologies can further support enhanced displays, such as improving contrast and clarity and scaling images. These features will effectively improve the experience of remote users.

\subsubsection{Enhancing Location Awareness Among Collaborators.}
We suggest that telepresence robots used for collaborative locomotion should provide remote users with information about their partner's location and movement status. Our study results show that enhanced positional awareness allows local users to walk more closely with the robot, aligning more closely with the behavior exhibited by partners in co-located collaboration, which is consistent with previous work~\cite{gong2023side}. The potential advantages of this information transfer include, first, enhancing the remote user's sense of presence and participation, similar to the experience of walking in a real-world space. Additionally, the awareness of the partner's position can alleviate the burden and psychological pressure of remote control. Support for positional awareness can effectively improve the working experience in remote collaboration.

\subsubsection{Providing More Ways for Emotional Communication.}
In this study, we used force sensors to simulate the intimate action of shoulder-tapping between collaborators in real life. Although participants mentioned that this touch interaction currently provides too little feedback to remote users, they still recognized that this interaction brought more fun and closeness to the task. Numerous studies have shown that touch has significant benefits for emotional interaction~\cite{guo2023touch,zhan2023enable}. We suggest that future robots designed for remote collaboration could offer more interactive ways of behaving to further enhance emotional communication between collaborators.

\subsubsection{Supporting Collaborative Partners in Jointly Referencing Environmental Information.}
~\ryrevised{In our research, we employed posture recognition algorithms and projection technology~\cite{li2023understanding} to streamline the exchange of referential information between remote and local participants. Our findings underscore the importance of improving the ability of both parties to reference effectively, which plays a pivotal role in facilitating successful collaborative locomotion. It is critical for local participants to accurately perceive the cues given by their remote counterparts, and vice versa, enhancing the remote participants' understanding of the local cues is equally vital. Considering the constrained visual capabilities of remote users, leveraging technology to better showcase the local participants' referential gestures can swiftly aid remote users in grasping and reacting to their partners' movements. By concentrating on bolstering the referential communication for both local and remote participants, rather than focusing exclusively on the needs of remote users, we can achieve a more efficient exchange of information and enhance the overall collaborative experience.}

\subsection{Limitations and Future Work}

\ryrevised{This study delves into dyadic collaborative exhibition viewing within the broader context of collaborative locomotion activities. Through observing exhibitions with varying levels of visitor traffic, we developed the "User-Partner-Environment" framework. This framework delineates the essential awareness needs and behaviors, thereby enabling the TeleAware robot to enhance support for collaborative locomotion. Nonetheless, in real-world settings, scenarios involving groups of three or more individuals are common. Investigating the framework's applicability to such group viewing scenarios and identifying how it might be expanded to accommodate a wider array of collaborative locomotion situations are areas ripe for further exploration. We hypothesize that the core awareness needs of the framework will not undergo significant changes, but the framework itself will become more multi-dimensional. Additionally, while we discussed the impact of visitor flow when deducing the "User-Partner-Environment" framework, the task environment design was purely based on a dual-user collaborative scenario, lacking further validation in crowded settings. Therefore, we plan to expand the number of involved users and further validate and improve the framework through in-depth observational studies.}


~\ryrevised{In addition to focusing on the overarching system design and introducing innovations, this study integrates various components to bolster multiple types of awareness, thereby enhancing the TeleAware system's effectiveness in two-person collaborative locomotion scenarios. While the TeleAware robot demonstrates the potential to foster social closeness and maintain mutual awareness among users, the current study does not pinpoint which specific components are most critical. Moving forward, we plan to undertake detailed experimental designs for each component to precisely evaluate their contributions to enhancing awareness.}


Finally, due to limited development time, some functions of TeleAware are still at the prototype verification stage, meaning that many features have room for further improvement. Currently, the referential projection only realizes ground projection beams and the direction of the beams does not always align with the direction the user intends to refer to, so the feature still has ambiguity. In the future, we plan to further expand the projection feature and explore its full potential, such as supporting the expression of different movement states of the robot through dynamic beams. Additionally, the shoulder-tapping feature's current automatic trigger rotation conflicts with the remote user's control of the robot. Based on the validation of its effectiveness in this study, we hope to provide remote users with more natural interaction metaphors and feedback through auditory or visual feedback in the future.

\section{Conclusion}
In this paper, we introduce the TeleAware robot, a telepresence robot system designed to support collaborative locomotion. ~\ryrevised{Through an observational study of colocated collaborative exhibition touring settings, we identified a User-Partner-Environment awareness design space and established a set of design goals aimed at enhancing awareness as a framework. Guided by these goals, we designed and developed awareness-augmenting features for the TeleAware Robot, aimed at supporting the experience in collaborative locomotion. Based on the results of simulated guided visiting tasks, the TeleAware Robot shows the potential to enhance lower workload, closer social proximity, higher mutual awareness, and social presence. We also discuss the impact of mobility and roles between local and remote users and propose design implications to guide the future design of telepresence robots in remote collaborative locomotion.}

\begin{acks}
This project is supported by National Natural Science Foundation Youth Fund 62202267. We extend our sincere gratitude to all the participants in our experiment.
\end{acks}

\bibliographystyle{ACM-Reference-Format}
\bibliography{main}


\begin{thebibliography}{81}


\ifx \showCODEN    \undefined \def \showCODEN     #1{\unskip}     \fi
\ifx \showDOI      \undefined \def \showDOI       #1{#1}\fi
\ifx \showISBNx    \undefined \def \showISBNx     #1{\unskip}     \fi
\ifx \showISBNxiii \undefined \def \showISBNxiii  #1{\unskip}     \fi
\ifx \showISSN     \undefined \def \showISSN      #1{\unskip}     \fi
\ifx \showLCCN     \undefined \def \showLCCN      #1{\unskip}     \fi
\ifx \shownote     \undefined \def \shownote      #1{#1}          \fi
\ifx \showarticletitle \undefined \def \showarticletitle #1{#1}   \fi
\ifx \showURL      \undefined \def \showURL       {\relax}        \fi
\providecommand\bibfield[2]{#2}
\providecommand\bibinfo[2]{#2}
\providecommand\natexlab[1]{#1}
\providecommand\showeprint[2][]{arXiv:#2}

\bibitem[Adalgeirsson and Breazeal(2010)]%
        {adalgeirsson2010mebot}
\bibfield{author}{\bibinfo{person}{Sigurdur~Orn Adalgeirsson} {and} \bibinfo{person}{Cynthia Breazeal}.} \bibinfo{year}{2010}\natexlab{}.
\newblock \showarticletitle{MeBot: A robotic platform for socially embodied telepresence}. In \bibinfo{booktitle}{\emph{2010 5th ACM/IEEE International Conference on Human-Robot Interaction (HRI)}}. IEEE, \bibinfo{pages}{15--22}.
\newblock


\bibitem[Alem and Huang(2011)]%
        {alem2011developing}
\bibfield{author}{\bibinfo{person}{Leila Alem} {and} \bibinfo{person}{Weidong Huang}.} \bibinfo{year}{2011}\natexlab{}.
\newblock \showarticletitle{Developing mobile remote collaboration systems for industrial use: some design challenges}. In \bibinfo{booktitle}{\emph{Human-Computer Interaction--INTERACT 2011: 13th IFIP TC 13 International Conference, Lisbon, Portugal, September 5-9, 2011, Proceedings, Part IV 13}}. Springer, \bibinfo{pages}{442--445}.
\newblock


\bibitem[Aron et~al\mbox{.}(1992)]%
        {aron1992inclusion}
\bibfield{author}{\bibinfo{person}{Arthur Aron}, \bibinfo{person}{Elaine~N Aron}, {and} \bibinfo{person}{Danny Smollan}.} \bibinfo{year}{1992}\natexlab{}.
\newblock \showarticletitle{Inclusion of other in the self scale and the structure of interpersonal closeness.}
\newblock \bibinfo{journal}{\emph{Journal of personality and social psychology}} \bibinfo{volume}{63}, \bibinfo{number}{4} (\bibinfo{year}{1992}), \bibinfo{pages}{596}.
\newblock


\bibitem[Bazzano et~al\mbox{.}(2017)]%
        {bazzano2017comparing}
\bibfield{author}{\bibinfo{person}{Federica Bazzano}, \bibinfo{person}{Fabrizio Lamberti}, \bibinfo{person}{Andrea Sanna}, \bibinfo{person}{Gianluca Paravati}, {and} \bibinfo{person}{Marco Gaspardone}.} \bibinfo{year}{2017}\natexlab{}.
\newblock \showarticletitle{Comparing Usability of User Interfaces for Robotic Telepresence.}. In \bibinfo{booktitle}{\emph{VISIGRAPP (2: HUCAPP)}}. \bibinfo{pages}{46--54}.
\newblock


\bibitem[Beraldo et~al\mbox{.}(2021)]%
        {beraldo2021shared}
\bibfield{author}{\bibinfo{person}{Gloria Beraldo}, \bibinfo{person}{Kenji Koide}, \bibinfo{person}{Amedeo Cesta}, \bibinfo{person}{Satoshi Hoshino}, \bibinfo{person}{Jun Miura}, \bibinfo{person}{Matteo Salv{\`a}}, {and} \bibinfo{person}{Emanuele Menegatti}.} \bibinfo{year}{2021}\natexlab{}.
\newblock \showarticletitle{Shared autonomy for telepresence robots based on people-aware navigation}. In \bibinfo{booktitle}{\emph{International Conference on Intelligent Autonomous Systems}}. Springer, \bibinfo{pages}{109--122}.
\newblock


\bibitem[Biocca et~al\mbox{.}(2001)]%
        {biocca2001networked}
\bibfield{author}{\bibinfo{person}{Frank Biocca}, \bibinfo{person}{Chad Harms}, {and} \bibinfo{person}{Jenn Gregg}.} \bibinfo{year}{2001}\natexlab{}.
\newblock \showarticletitle{The networked minds measure of social presence: Pilot test of the factor structure and concurrent validity}. In \bibinfo{booktitle}{\emph{4th annual international workshop on presence, Philadelphia, PA}}. \bibinfo{pages}{1--9}.
\newblock


\bibitem[Botev and Rodr{\'\i}guez~Lera(2020)]%
        {botev2020immersive}
\bibfield{author}{\bibinfo{person}{Jean Botev} {and} \bibinfo{person}{Francisco~J Rodr{\'\i}guez~Lera}.} \bibinfo{year}{2020}\natexlab{}.
\newblock \showarticletitle{Immersive telepresence framework for remote educational scenarios}. In \bibinfo{booktitle}{\emph{Learning and Collaboration Technologies. Human and Technology Ecosystems: 7th International Conference, LCT 2020, Held as Part of the 22nd HCI International Conference, HCII 2020, Copenhagen, Denmark, July 19--24, 2020, Proceedings, Part II 22}}. Springer, \bibinfo{pages}{373--390}.
\newblock


\bibitem[Burgard et~al\mbox{.}(2003)]%
        {burgard2003tele}
\bibfield{author}{\bibinfo{person}{Wolfram Burgard}, \bibinfo{person}{Panos Trahanias}, \bibinfo{person}{Dirk H{\"a}hnel}, \bibinfo{person}{Mark Moors}, \bibinfo{person}{Dirk Schulz}, \bibinfo{person}{Haris Baltzakis}, {and} \bibinfo{person}{Antonis Argyros}.} \bibinfo{year}{2003}\natexlab{}.
\newblock \showarticletitle{Tele-presence in populated exhibitions through web-operated mobile robots}.
\newblock \bibinfo{journal}{\emph{Autonomous Robots}}  \bibinfo{volume}{15} (\bibinfo{year}{2003}), \bibinfo{pages}{299--316}.
\newblock


\bibitem[Ching et~al\mbox{.}(2016)]%
        {ching2016design}
\bibfield{author}{\bibinfo{person}{Pang~Wee Ching}, \bibinfo{person}{Wong~Choon Yue}, {and} \bibinfo{person}{Gerald Seet~Gim Lee}.} \bibinfo{year}{2016}\natexlab{}.
\newblock \showarticletitle{Design and development of edgar-a telepresence humanoid for robot-mediated communication and social applications}. In \bibinfo{booktitle}{\emph{2016 IEEE International Conference on Control and Robotics Engineering (ICCRE)}}. IEEE, \bibinfo{pages}{1--4}.
\newblock


\bibitem[Chu et~al\mbox{.}(2023)]%
        {chu2023work}
\bibfield{author}{\bibinfo{person}{Mengdi Chu}, \bibinfo{person}{Keyu Zong}, \bibinfo{person}{Xin Shu}, \bibinfo{person}{Jiangtao Gong}, \bibinfo{person}{Zhicong Lu}, \bibinfo{person}{Kaimin Guo}, \bibinfo{person}{Xinyi Dai}, {and} \bibinfo{person}{Guyue Zhou}.} \bibinfo{year}{2023}\natexlab{}.
\newblock \showarticletitle{Work with AI and Work for AI: Autonomous Vehicle Safety Drivers’ Lived Experiences}. In \bibinfo{booktitle}{\emph{Proceedings of the 2023 CHI Conference on Human Factors in Computing Systems}}. \bibinfo{pages}{1--16}.
\newblock


\bibitem[Couzin et~al\mbox{.}(2003)]%
        {couzin2003self}
\bibfield{author}{\bibinfo{person}{Iain~D Couzin}, \bibinfo{person}{Jens Krause}, {et~al\mbox{.}}} \bibinfo{year}{2003}\natexlab{}.
\newblock \showarticletitle{Self-organization and collective behavior in vertebrates}.
\newblock \bibinfo{journal}{\emph{Advances in the Study of Behavior}} \bibinfo{volume}{32}, \bibinfo{number}{1} (\bibinfo{year}{2003}), \bibinfo{pages}{10--1016}.
\newblock


\bibitem[Dillenbourg and Traum(1997)]%
        {dillenbourg1997role}
\bibfield{author}{\bibinfo{person}{P Dillenbourg} {and} \bibinfo{person}{D Traum}.} \bibinfo{year}{1997}\natexlab{}.
\newblock \showarticletitle{The role of a whiteboard in a distributed cognitive system}. In \bibinfo{booktitle}{\emph{Swiss Workshop on Distributed and Collaborative Systems, Lausanne, Switzerland}}.
\newblock


\bibitem[Dim and Kuflik(2014)]%
        {dim2014automatic}
\bibfield{author}{\bibinfo{person}{Eyal Dim} {and} \bibinfo{person}{Tsvi Kuflik}.} \bibinfo{year}{2014}\natexlab{}.
\newblock \showarticletitle{Automatic detection of social behavior of museum visitor pairs}.
\newblock \bibinfo{journal}{\emph{ACM Transactions on Interactive Intelligent Systems (TiiS)}} \bibinfo{volume}{4}, \bibinfo{number}{4} (\bibinfo{year}{2014}), \bibinfo{pages}{1--30}.
\newblock


\bibitem[Dourish and Bellotti(1992)]%
        {dourish1992awareness}
\bibfield{author}{\bibinfo{person}{Paul Dourish} {and} \bibinfo{person}{Victoria Bellotti}.} \bibinfo{year}{1992}\natexlab{}.
\newblock \showarticletitle{Awareness and coordination in shared workspaces}. In \bibinfo{booktitle}{\emph{Proceedings of the 1992 ACM conference on Computer-supported cooperative work}}. \bibinfo{pages}{107--114}.
\newblock


\bibitem[Evrard and Kheddar(2009)]%
        {evrard2009homotopy}
\bibfield{author}{\bibinfo{person}{Paul Evrard} {and} \bibinfo{person}{Abderrahmane Kheddar}.} \bibinfo{year}{2009}\natexlab{}.
\newblock \showarticletitle{Homotopy switching model for dyad haptic interaction in physical collaborative tasks}. In \bibinfo{booktitle}{\emph{World haptics 2009-third joint EuroHaptics conference and symposium on haptic interfaces for virtual environment and teleoperator systems}}. IEEE, \bibinfo{pages}{45--50}.
\newblock


\bibitem[Fink et~al\mbox{.}(2022)]%
        {fink2022re}
\bibfield{author}{\bibinfo{person}{Daniel~Immanuel Fink}, \bibinfo{person}{Johannes Zagermann}, \bibinfo{person}{Harald Reiterer}, {and} \bibinfo{person}{Hans-Christian Jetter}.} \bibinfo{year}{2022}\natexlab{}.
\newblock \showarticletitle{Re-locations: Augmenting Personal and Shared Workspaces to Support Remote Collaboration in Incongruent Spaces}.
\newblock \bibinfo{journal}{\emph{Proceedings of the ACM on Human-Computer Interaction}} \bibinfo{volume}{6}, \bibinfo{number}{ISS} (\bibinfo{year}{2022}), \bibinfo{pages}{1--30}.
\newblock


\bibitem[Fischedick et~al\mbox{.}(2023)]%
        {fischedick2023bridging}
\bibfield{author}{\bibinfo{person}{Soehnke~Benedikt Fischedick}, \bibinfo{person}{Kay Richter}, \bibinfo{person}{Tim Wengefeld}, \bibinfo{person}{Daniel Seichter}, \bibinfo{person}{Andrea Scheidig}, \bibinfo{person}{Nicola Doering}, \bibinfo{person}{Wolfgang Broll}, \bibinfo{person}{Stephan Werner}, \bibinfo{person}{Alexander Raake}, {and} \bibinfo{person}{Horst-Michael Gross}.} \bibinfo{year}{2023}\natexlab{}.
\newblock \showarticletitle{Bridging Distance with a Collaborative Telepresence Robot for Older Adults--Report on Progress in the CO-HUMANICS Project}. In \bibinfo{booktitle}{\emph{ISR Europe 2023; 56th International Symposium on Robotics}}. VDE, \bibinfo{pages}{346--353}.
\newblock


\bibitem[Fransen et~al\mbox{.}(2011)]%
        {fransen2011mediating}
\bibfield{author}{\bibinfo{person}{Jos Fransen}, \bibinfo{person}{Paul~A Kirschner}, {and} \bibinfo{person}{Gijsbert Erkens}.} \bibinfo{year}{2011}\natexlab{}.
\newblock \showarticletitle{Mediating team effectiveness in the context of collaborative learning: The importance of team and task awareness}.
\newblock \bibinfo{journal}{\emph{Computers in human Behavior}} \bibinfo{volume}{27}, \bibinfo{number}{3} (\bibinfo{year}{2011}), \bibinfo{pages}{1103--1113}.
\newblock


\bibitem[Gao et~al\mbox{.}(2023)]%
        {gao2023agent}
\bibfield{author}{\bibinfo{person}{Qi Gao}, \bibinfo{person}{Wei Xu}, \bibinfo{person}{Mowei Shen}, {and} \bibinfo{person}{Zaifeng Gao}.} \bibinfo{year}{2023}\natexlab{}.
\newblock \showarticletitle{Agent Teaming Situation Awareness (ATSA): A Situation Awareness Framework for Human-AI Teaming}.
\newblock \bibinfo{journal}{\emph{arXiv preprint arXiv:2308.16785}} (\bibinfo{year}{2023}).
\newblock


\bibitem[Gong et~al\mbox{.}(2021)]%
        {gong2021holoboard}
\bibfield{author}{\bibinfo{person}{Jiangtao Gong}, \bibinfo{person}{Teng Han}, \bibinfo{person}{Siling Guo}, \bibinfo{person}{Jiannan Li}, \bibinfo{person}{Siyu Zha}, \bibinfo{person}{Liuxin Zhang}, \bibinfo{person}{Feng Tian}, \bibinfo{person}{Qianying Wang}, {and} \bibinfo{person}{Yong Rui}.} \bibinfo{year}{2021}\natexlab{}.
\newblock \showarticletitle{Holoboard: A large-format immersive teaching board based on pseudo holographics}. In \bibinfo{booktitle}{\emph{The 34th Annual ACM Symposium on User Interface Software and Technology}}. \bibinfo{pages}{441--456}.
\newblock


\bibitem[Gong et~al\mbox{.}(2023)]%
        {gong2023side}
\bibfield{author}{\bibinfo{person}{Jiangtao Gong}, \bibinfo{person}{Jingjing Sun}, \bibinfo{person}{Mengdi Chu}, \bibinfo{person}{Xiaoye Wang}, \bibinfo{person}{Minghao Luo}, \bibinfo{person}{Yi Lu}, \bibinfo{person}{Liuxin Zhang}, \bibinfo{person}{Yaqiang Wu}, \bibinfo{person}{Qianying Wang}, {and} \bibinfo{person}{Can Liu}.} \bibinfo{year}{2023}\natexlab{}.
\newblock \showarticletitle{Side-by-Side vs Face-to-Face: Evaluating Colocated Collaboration via a Transparent Wall-sized Display}.
\newblock \bibinfo{journal}{\emph{Proceedings of the ACM on Human-Computer Interaction}} \bibinfo{volume}{7}, \bibinfo{number}{CSCW1} (\bibinfo{year}{2023}), \bibinfo{pages}{1--29}.
\newblock


\bibitem[Guo et~al\mbox{.}(2019)]%
        {guo2019blocks}
\bibfield{author}{\bibinfo{person}{Anhong Guo}, \bibinfo{person}{Ilter Canberk}, \bibinfo{person}{Hannah Murphy}, \bibinfo{person}{Andr{\'e}s Monroy-Hern{\'a}ndez}, {and} \bibinfo{person}{Rajan Vaish}.} \bibinfo{year}{2019}\natexlab{}.
\newblock \showarticletitle{Blocks: Collaborative and persistent augmented reality experiences}.
\newblock \bibinfo{journal}{\emph{Proceedings of the ACM on Interactive, Mobile, Wearable and Ubiquitous Technologies}} \bibinfo{volume}{3}, \bibinfo{number}{3} (\bibinfo{year}{2019}), \bibinfo{pages}{1--24}.
\newblock


\bibitem[Guo et~al\mbox{.}(2023)]%
        {guo2023touch}
\bibfield{author}{\bibinfo{person}{Shihui Guo}, \bibinfo{person}{Lishuang Zhan}, \bibinfo{person}{Yancheng Cao}, \bibinfo{person}{Chen Zheng}, \bibinfo{person}{Guyue Zhou}, {and} \bibinfo{person}{Jiangtao Gong}.} \bibinfo{year}{2023}\natexlab{}.
\newblock \showarticletitle{Touch-and-Heal: Data-driven Affective Computing in Tactile Interaction with Robotic Dog}.
\newblock \bibinfo{journal}{\emph{Proceedings of the ACM on Interactive, Mobile, Wearable and Ubiquitous Technologies}} \bibinfo{volume}{7}, \bibinfo{number}{2} (\bibinfo{year}{2023}), \bibinfo{pages}{1--33}.
\newblock


\bibitem[Hart(2006)]%
        {hart2006nasa}
\bibfield{author}{\bibinfo{person}{Sandra~G Hart}.} \bibinfo{year}{2006}\natexlab{}.
\newblock \showarticletitle{NASA-task load index (NASA-TLX); 20 years later}. In \bibinfo{booktitle}{\emph{Proceedings of the human factors and ergonomics society annual meeting}}, Vol.~\bibinfo{volume}{50}. Sage publications Sage CA: Los Angeles, CA, \bibinfo{pages}{904--908}.
\newblock


\bibitem[Hasegawa and Nakauchi(2014)]%
        {hasegawa2014telepresence}
\bibfield{author}{\bibinfo{person}{Komei Hasegawa} {and} \bibinfo{person}{Yasushi Nakauchi}.} \bibinfo{year}{2014}\natexlab{}.
\newblock \showarticletitle{Telepresence robot that exaggerates non-verbal cues for taking turns in multi-party teleconferences}. In \bibinfo{booktitle}{\emph{Proceedings of the second international conference on Human-agent interaction}}. \bibinfo{pages}{293--296}.
\newblock


\bibitem[Helbing et~al\mbox{.}(2001)]%
        {helbing2001self}
\bibfield{author}{\bibinfo{person}{Dirk Helbing}, \bibinfo{person}{P{\'e}ter Moln{\'a}r}, \bibinfo{person}{Ill{\'e}s~J Farkas}, {and} \bibinfo{person}{Kai Bolay}.} \bibinfo{year}{2001}\natexlab{}.
\newblock \showarticletitle{Self-organizing pedestrian movement}.
\newblock \bibinfo{journal}{\emph{Environment and planning B: planning and design}} \bibinfo{volume}{28}, \bibinfo{number}{3} (\bibinfo{year}{2001}), \bibinfo{pages}{361--383}.
\newblock


\bibitem[Herbort and Kunde(2016)]%
        {herbort2016spatial}
\bibfield{author}{\bibinfo{person}{Oliver Herbort} {and} \bibinfo{person}{Wilfried Kunde}.} \bibinfo{year}{2016}\natexlab{}.
\newblock \showarticletitle{Spatial (mis-) interpretation of pointing gestures to distal referents.}
\newblock \bibinfo{journal}{\emph{Journal of Experimental Psychology: Human Perception and Performance}} \bibinfo{volume}{42}, \bibinfo{number}{1} (\bibinfo{year}{2016}), \bibinfo{pages}{78}.
\newblock


\bibitem[Herbort and Kunde(2018)]%
        {herbort2018point}
\bibfield{author}{\bibinfo{person}{Oliver Herbort} {and} \bibinfo{person}{Wilfried Kunde}.} \bibinfo{year}{2018}\natexlab{}.
\newblock \showarticletitle{How to point and to interpret pointing gestures? Instructions can reduce pointer--observer misunderstandings}.
\newblock \bibinfo{journal}{\emph{Psychological research}} \bibinfo{volume}{82}, \bibinfo{number}{2} (\bibinfo{year}{2018}), \bibinfo{pages}{395--406}.
\newblock


\bibitem[Heshmat et~al\mbox{.}(2018)]%
        {heshmat2018geocaching}
\bibfield{author}{\bibinfo{person}{Yasamin Heshmat}, \bibinfo{person}{Brennan Jones}, \bibinfo{person}{Xiaoxuan Xiong}, \bibinfo{person}{Carman Neustaedter}, \bibinfo{person}{Anthony Tang}, \bibinfo{person}{Bernhard~E Riecke}, {and} \bibinfo{person}{Lillian Yang}.} \bibinfo{year}{2018}\natexlab{}.
\newblock \showarticletitle{Geocaching with a beam: Shared outdoor activities through a telepresence robot with 360 degree viewing}. In \bibinfo{booktitle}{\emph{Proceedings of the 2018 CHI Conference on Human Factors in Computing Systems}}. \bibinfo{pages}{1--13}.
\newblock


\bibitem[Hetherington et~al\mbox{.}(2021)]%
        {hetherington2021hey}
\bibfield{author}{\bibinfo{person}{Nicholas~J Hetherington}, \bibinfo{person}{Elizabeth~A Croft}, {and} \bibinfo{person}{HF~Machiel Van~der Loos}.} \bibinfo{year}{2021}\natexlab{}.
\newblock \showarticletitle{Hey robot, which way are you going? nonverbal motion legibility cues for human-robot spatial interaction}.
\newblock \bibinfo{journal}{\emph{IEEE Robotics and Automation Letters}} \bibinfo{volume}{6}, \bibinfo{number}{3} (\bibinfo{year}{2021}), \bibinfo{pages}{5010--5015}.
\newblock


\bibitem[Hove and Risen(2009)]%
        {hove2009s}
\bibfield{author}{\bibinfo{person}{Michael~J Hove} {and} \bibinfo{person}{Jane~L Risen}.} \bibinfo{year}{2009}\natexlab{}.
\newblock \showarticletitle{It's all in the timing: Interpersonal synchrony increases affiliation}.
\newblock \bibinfo{journal}{\emph{Social cognition}} \bibinfo{volume}{27}, \bibinfo{number}{6} (\bibinfo{year}{2009}), \bibinfo{pages}{949--960}.
\newblock


\bibitem[Ishikawa(2021)]%
        {ishikawa2021spatial}
\bibfield{author}{\bibinfo{person}{Toru Ishikawa}.} \bibinfo{year}{2021}\natexlab{}.
\newblock \showarticletitle{Spatial thinking, cognitive mapping, and spatial awareness}.
\newblock \bibinfo{journal}{\emph{Cognitive Processing}} \bibinfo{volume}{22}, \bibinfo{number}{Suppl 1} (\bibinfo{year}{2021}), \bibinfo{pages}{89--96}.
\newblock


\bibitem[Jia et~al\mbox{.}(2021)]%
        {jia2021tiui}
\bibfield{author}{\bibinfo{person}{Yunde Jia}, \bibinfo{person}{Yanmei Dong}, \bibinfo{person}{Bin Xu}, {and} \bibinfo{person}{Che Sun}.} \bibinfo{year}{2021}\natexlab{}.
\newblock \showarticletitle{Tiui: Touching live video for telepresence operation}.
\newblock \bibinfo{journal}{\emph{IEEE Transactions on Mobile Computing}} (\bibinfo{year}{2021}).
\newblock


\bibitem[Jones et~al\mbox{.}(2021)]%
        {jones2021belonging}
\bibfield{author}{\bibinfo{person}{Brennan Jones}, \bibinfo{person}{Yaying Zhang}, \bibinfo{person}{Priscilla~NY Wong}, {and} \bibinfo{person}{Sean Rintel}.} \bibinfo{year}{2021}\natexlab{}.
\newblock \showarticletitle{Belonging there: VROOM-ing into the uncanny valley of XR telepresence}.
\newblock \bibinfo{journal}{\emph{Proceedings of the ACM on Human-Computer Interaction}} \bibinfo{volume}{5}, \bibinfo{number}{CSCW1} (\bibinfo{year}{2021}), \bibinfo{pages}{1--31}.
\newblock


\bibitem[King(2010)]%
        {king2010follow}
\bibfield{author}{\bibinfo{person}{Andrew~J King}.} \bibinfo{year}{2010}\natexlab{}.
\newblock \showarticletitle{Follow me! I’ma leader if you do; I’ma failed initiator if you don’t}.
\newblock \bibinfo{journal}{\emph{Behavioural processes}} \bibinfo{volume}{84}, \bibinfo{number}{3} (\bibinfo{year}{2010}), \bibinfo{pages}{671--674}.
\newblock


\bibitem[Krause et~al\mbox{.}(2000)]%
        {krause2000leadership}
\bibfield{author}{\bibinfo{person}{J Krause}, \bibinfo{person}{D Hoare}, \bibinfo{person}{S Krause}, \bibinfo{person}{CK Hemelrijk}, {and} \bibinfo{person}{DI Rubenstein}.} \bibinfo{year}{2000}\natexlab{}.
\newblock \showarticletitle{Leadership in fish shoals}.
\newblock \bibinfo{journal}{\emph{Fish and Fisheries}} \bibinfo{volume}{1}, \bibinfo{number}{1} (\bibinfo{year}{2000}), \bibinfo{pages}{82--89}.
\newblock


\bibitem[Kulyk et~al\mbox{.}(2008)]%
        {kulyk2008situational}
\bibfield{author}{\bibinfo{person}{Olga Kulyk}, \bibinfo{person}{Gerrit van~der Veer}, {and} \bibinfo{person}{Betsy van Dijk}.} \bibinfo{year}{2008}\natexlab{}.
\newblock \showarticletitle{Situational awareness support to enhance teamwork in collaborative environments}. In \bibinfo{booktitle}{\emph{Proceedings of the 15th European conference on Cognitive ergonomics: the ergonomics of cool interaction}}. \bibinfo{pages}{1--5}.
\newblock


\bibitem[Lambropoulos et~al\mbox{.}(2012)]%
        {lambropoulos2012supporting}
\bibfield{author}{\bibinfo{person}{Niki Lambropoulos}, \bibinfo{person}{Xristine Faulkner}, {and} \bibinfo{person}{Fintan Culwin}.} \bibinfo{year}{2012}\natexlab{}.
\newblock \showarticletitle{Supporting social awareness in collaborative e-learning}.
\newblock \bibinfo{journal}{\emph{British Journal of Educational Technology}} \bibinfo{volume}{43}, \bibinfo{number}{2} (\bibinfo{year}{2012}), \bibinfo{pages}{295--306}.
\newblock


\bibitem[Lamme(2003)]%
        {lamme2003visual}
\bibfield{author}{\bibinfo{person}{Victor~AF Lamme}.} \bibinfo{year}{2003}\natexlab{}.
\newblock \showarticletitle{Why visual attention and awareness are different}.
\newblock \bibinfo{journal}{\emph{Trends in cognitive sciences}} \bibinfo{volume}{7}, \bibinfo{number}{1} (\bibinfo{year}{2003}), \bibinfo{pages}{12--18}.
\newblock


\bibitem[Launay et~al\mbox{.}(2013)]%
        {launay2013synchronization}
\bibfield{author}{\bibinfo{person}{Jacques Launay}, \bibinfo{person}{Roger~T Dean}, {and} \bibinfo{person}{Freya Bailes}.} \bibinfo{year}{2013}\natexlab{}.
\newblock \showarticletitle{Synchronization can influence trust following virtual interaction}.
\newblock \bibinfo{journal}{\emph{Experimental psychology}} (\bibinfo{year}{2013}).
\newblock


\bibitem[Li et~al\mbox{.}(2019)]%
        {li2019communicating}
\bibfield{author}{\bibinfo{person}{Jamy Li}, \bibinfo{person}{Andrea Cuadra}, \bibinfo{person}{Brian Mok}, \bibinfo{person}{Byron Reeves}, \bibinfo{person}{Jofish Kaye}, {and} \bibinfo{person}{Wendy Ju}.} \bibinfo{year}{2019}\natexlab{}.
\newblock \showarticletitle{Communicating dominance in a nonanthropomorphic robot using locomotion}.
\newblock \bibinfo{journal}{\emph{ACM Transactions on Human-Robot Interaction (THRI)}} \bibinfo{volume}{8}, \bibinfo{number}{1} (\bibinfo{year}{2019}), \bibinfo{pages}{1--14}.
\newblock


\bibitem[Li et~al\mbox{.}(2023)]%
        {li2023understanding}
\bibfield{author}{\bibinfo{person}{Yang Li}, \bibinfo{person}{Xiaoxue Chen}, \bibinfo{person}{Hao Zhao}, \bibinfo{person}{Jiangtao Gong}, \bibinfo{person}{Guyue Zhou}, \bibinfo{person}{Federico Rossano}, {and} \bibinfo{person}{Yixin Zhu}.} \bibinfo{year}{2023}\natexlab{}.
\newblock \showarticletitle{Understanding Embodied Reference with Touch-Line Transformer.}. In \bibinfo{booktitle}{\emph{ICLR}}.
\newblock


\bibitem[Lu et~al\mbox{.}(2022)]%
        {lu2022classification}
\bibfield{author}{\bibinfo{person}{Yi Lu}, \bibinfo{person}{Xiaoye Wang}, \bibinfo{person}{Jiangtao Gong}, \bibinfo{person}{Lejia Zhou}, \bibinfo{person}{Sen Ge}, {et~al\mbox{.}}} \bibinfo{year}{2022}\natexlab{}.
\newblock \showarticletitle{Classification, application, challenge, and future of midair gestures in augmented reality}.
\newblock \bibinfo{journal}{\emph{Journal of Sensors}}  \bibinfo{volume}{2022} (\bibinfo{year}{2022}).
\newblock


\bibitem[Machino et~al\mbox{.}(2006)]%
        {machino2006remote}
\bibfield{author}{\bibinfo{person}{Tamotsu Machino}, \bibinfo{person}{Satoshi Iwaki}, \bibinfo{person}{Hiroaki Kawata}, \bibinfo{person}{Yoshimasa Yanagihara}, \bibinfo{person}{Yoshito Nanjo}, {and} \bibinfo{person}{K-i Shimokura}.} \bibinfo{year}{2006}\natexlab{}.
\newblock \showarticletitle{Remote-collaboration system using mobile robot with camera and projector}. In \bibinfo{booktitle}{\emph{Proceedings 2006 IEEE International Conference on Robotics and Automation, 2006. ICRA 2006.}} IEEE, \bibinfo{pages}{4063--4068}.
\newblock


\bibitem[Matsuda and Rekimoto(2016)]%
        {matsuda2016scalablebody}
\bibfield{author}{\bibinfo{person}{Akira Matsuda} {and} \bibinfo{person}{Jun Rekimoto}.} \bibinfo{year}{2016}\natexlab{}.
\newblock \showarticletitle{Scalablebody: A telepresence robot supporting socially acceptable interactions and human augmentation through vertical actuation}. In \bibinfo{booktitle}{\emph{Adjunct Proceedings of the 29th Annual ACM Symposium on User Interface Software and Technology}}. \bibinfo{pages}{103--105}.
\newblock


\bibitem[Neustaedter et~al\mbox{.}(2016)]%
        {neustaedter2016beam}
\bibfield{author}{\bibinfo{person}{Carman Neustaedter}, \bibinfo{person}{Gina Venolia}, \bibinfo{person}{Jason Procyk}, {and} \bibinfo{person}{Daniel Hawkins}.} \bibinfo{year}{2016}\natexlab{}.
\newblock \showarticletitle{To Beam or not to Beam: A study of remote telepresence attendance at an academic conference}. In \bibinfo{booktitle}{\emph{Proceedings of the 19th acm conference on computer-supported cooperative work \& social computing}}. \bibinfo{pages}{418--431}.
\newblock


\bibitem[Nova et~al\mbox{.}(2007)]%
        {nova2007collaboration}
\bibfield{author}{\bibinfo{person}{N Nova}, \bibinfo{person}{T Wehrle}, \bibinfo{person}{J Goslin}, \bibinfo{person}{Y Bourquin}, {and} \bibinfo{person}{P Dillenbourg}.} \bibinfo{year}{2007}\natexlab{}.
\newblock \bibinfo{booktitle}{\emph{Collaboration in a video game: Impacts of location awareness}}.
\newblock \bibinfo{type}{{T}echnical {R}eport}.
\newblock


\bibitem[Onishi et~al\mbox{.}(2016)]%
        {onishi2016embodiment}
\bibfield{author}{\bibinfo{person}{Yuya Onishi}, \bibinfo{person}{Kazuaki Tanaka}, {and} \bibinfo{person}{Hideyuki Nakanishi}.} \bibinfo{year}{2016}\natexlab{}.
\newblock \showarticletitle{Embodiment of video-mediated communication enhances social telepresence}. In \bibinfo{booktitle}{\emph{Proceedings of the Fourth International Conference on Human Agent Interaction}}. \bibinfo{pages}{171--178}.
\newblock


\bibitem[Orts-Escolano et~al\mbox{.}(2016)]%
        {orts2016holoportation}
\bibfield{author}{\bibinfo{person}{Sergio Orts-Escolano}, \bibinfo{person}{Christoph Rhemann}, \bibinfo{person}{Sean Fanello}, \bibinfo{person}{Wayne Chang}, \bibinfo{person}{Adarsh Kowdle}, \bibinfo{person}{Yury Degtyarev}, \bibinfo{person}{David Kim}, \bibinfo{person}{Philip~L Davidson}, \bibinfo{person}{Sameh Khamis}, \bibinfo{person}{Mingsong Dou}, {et~al\mbox{.}}} \bibinfo{year}{2016}\natexlab{}.
\newblock \showarticletitle{Holoportation: Virtual 3d teleportation in real-time}. In \bibinfo{booktitle}{\emph{Proceedings of the 29th annual symposium on user interface software and technology}}. \bibinfo{pages}{741--754}.
\newblock


\bibitem[Pejoska-Laajola et~al\mbox{.}(2017)]%
        {pejoska2017mobile}
\bibfield{author}{\bibinfo{person}{Jana Pejoska-Laajola}, \bibinfo{person}{Sanna Reponen}, \bibinfo{person}{Marjo Virnes}, {and} \bibinfo{person}{Teemu Leinonen}.} \bibinfo{year}{2017}\natexlab{}.
\newblock \showarticletitle{Mobile augmented communication for remote collaboration in a physical work context}.
\newblock \bibinfo{journal}{\emph{Australasian Journal of Educational Technology}} \bibinfo{volume}{33}, \bibinfo{number}{6} (\bibinfo{year}{2017}).
\newblock


\bibitem[Piumsomboon et~al\mbox{.}(2017)]%
        {piumsomboon2017covar}
\bibfield{author}{\bibinfo{person}{Thammathip Piumsomboon}, \bibinfo{person}{Youngho Lee}, \bibinfo{person}{Gun Lee}, {and} \bibinfo{person}{Mark Billinghurst}.} \bibinfo{year}{2017}\natexlab{}.
\newblock \showarticletitle{CoVAR: a collaborative virtual and augmented reality system for remote collaboration}.
\newblock In \bibinfo{booktitle}{\emph{SIGGRAPH Asia 2017 Emerging Technologies}}. \bibinfo{pages}{1--2}.
\newblock


\bibitem[Prajod et~al\mbox{.}(2023)]%
        {prajod2023gaze}
\bibfield{author}{\bibinfo{person}{Pooja Prajod}, \bibinfo{person}{Matteo Lavit~Nicora}, \bibinfo{person}{Matteo Malosio}, {and} \bibinfo{person}{Elisabeth Andr{\'e}}.} \bibinfo{year}{2023}\natexlab{}.
\newblock \showarticletitle{Gaze-based Attention Recognition for Human-Robot Collaboration}. In \bibinfo{booktitle}{\emph{Proceedings of the 16th International Conference on PErvasive Technologies Related to Assistive Environments}}. \bibinfo{pages}{140--147}.
\newblock


\bibitem[Rae et~al\mbox{.}(2013)]%
        {rae2013body}
\bibfield{author}{\bibinfo{person}{Irene Rae}, \bibinfo{person}{Leila Takayama}, {and} \bibinfo{person}{Bilge Mutlu}.} \bibinfo{year}{2013}\natexlab{}.
\newblock \showarticletitle{In-body experiences: embodiment, control, and trust in robot-mediated communication}. In \bibinfo{booktitle}{\emph{Proceedings of the SIGCHI Conference on Human Factors in Computing Systems}}. \bibinfo{pages}{1921--1930}.
\newblock


\bibitem[Rasch et~al\mbox{.}(2023)]%
        {rasch2023going}
\bibfield{author}{\bibinfo{person}{Julian Rasch}, \bibinfo{person}{Vladislav~Dmitrievic Rusakov}, \bibinfo{person}{Martin Schmitz}, {and} \bibinfo{person}{Florian M{\"u}ller}.} \bibinfo{year}{2023}\natexlab{}.
\newblock \showarticletitle{Going, Going, Gone: Exploring Intention Communication for Multi-User Locomotion in Virtual Reality}. In \bibinfo{booktitle}{\emph{Proceedings of the 2023 CHI Conference on Human Factors in Computing Systems}}. \bibinfo{pages}{1--13}.
\newblock


\bibitem[Reddish et~al\mbox{.}(2013)]%
        {reddish2013let}
\bibfield{author}{\bibinfo{person}{Paul Reddish}, \bibinfo{person}{Ronald Fischer}, {and} \bibinfo{person}{Joseph Bulbulia}.} \bibinfo{year}{2013}\natexlab{}.
\newblock \showarticletitle{Let’s dance together: Synchrony, shared intentionality and cooperation}.
\newblock \bibinfo{journal}{\emph{PloS one}} \bibinfo{volume}{8}, \bibinfo{number}{8} (\bibinfo{year}{2013}), \bibinfo{pages}{e71182}.
\newblock


\bibitem[Reed and Peshkin(2008)]%
        {reed2008physical}
\bibfield{author}{\bibinfo{person}{Kyle~B Reed} {and} \bibinfo{person}{Michael~A Peshkin}.} \bibinfo{year}{2008}\natexlab{}.
\newblock \showarticletitle{Physical collaboration of human-human and human-robot teams}.
\newblock \bibinfo{journal}{\emph{IEEE transactions on haptics}} \bibinfo{volume}{1}, \bibinfo{number}{2} (\bibinfo{year}{2008}), \bibinfo{pages}{108--120}.
\newblock


\bibitem[Rosenberg-Kima et~al\mbox{.}(2019)]%
        {rosenberg2019human}
\bibfield{author}{\bibinfo{person}{Rinat Rosenberg-Kima}, \bibinfo{person}{Yaacov Koren}, \bibinfo{person}{Maya Yachini}, {and} \bibinfo{person}{Goren Gordon}.} \bibinfo{year}{2019}\natexlab{}.
\newblock \showarticletitle{Human-Robot-Collaboration (HRC): social robots as teaching assistants for training activities in small groups}. In \bibinfo{booktitle}{\emph{2019 14th ACM/IEEE International Conference on Human-Robot Interaction (HRI)}}. IEEE, \bibinfo{pages}{522--523}.
\newblock


\bibitem[Ruddle et~al\mbox{.}(2015)]%
        {ruddle2015performance}
\bibfield{author}{\bibinfo{person}{Roy~A Ruddle}, \bibinfo{person}{Rhys~G Thomas}, \bibinfo{person}{Rebecca~S Randell}, \bibinfo{person}{Phil Quirke}, {and} \bibinfo{person}{Darren Treanor}.} \bibinfo{year}{2015}\natexlab{}.
\newblock \showarticletitle{Performance and interaction behaviour during visual search on large, high-resolution displays}.
\newblock \bibinfo{journal}{\emph{Information Visualization}} \bibinfo{volume}{14}, \bibinfo{number}{2} (\bibinfo{year}{2015}), \bibinfo{pages}{137--147}.
\newblock


\bibitem[Ruiz-del Solar et~al\mbox{.}(2021)]%
        {ruiz2021mental}
\bibfield{author}{\bibinfo{person}{Javier Ruiz-del Solar}, \bibinfo{person}{Mauricio Salazar}, \bibinfo{person}{Veronica Vargas-Araya}, \bibinfo{person}{Ulises Campodonico}, \bibinfo{person}{Nicolas Marticorena}, \bibinfo{person}{Giovanni Pais}, \bibinfo{person}{Rodrigo Salas}, \bibinfo{person}{Pablo Alfessi}, \bibinfo{person}{Victor~Contreras Rojas}, {and} \bibinfo{person}{Javier Urrutia}.} \bibinfo{year}{2021}\natexlab{}.
\newblock \showarticletitle{Mental and emotional health care for COVID-19 patients: employing Pudu, a telepresence robot}.
\newblock \bibinfo{journal}{\emph{IEEE Robotics \& Automation Magazine}} \bibinfo{volume}{28}, \bibinfo{number}{1} (\bibinfo{year}{2021}), \bibinfo{pages}{82--89}.
\newblock


\bibitem[Saadatian et~al\mbox{.}(2013)]%
        {saadatian2013personalizable}
\bibfield{author}{\bibinfo{person}{Elham Saadatian}, \bibinfo{person}{Hooman Samani}, \bibinfo{person}{Anshul Vikram}, \bibinfo{person}{Rahul Parsani}, \bibinfo{person}{Lenis~Tejada Rodriguez}, {and} \bibinfo{person}{Ryohei Nakatsu}.} \bibinfo{year}{2013}\natexlab{}.
\newblock \showarticletitle{Personalizable embodied telepresence system for remote interpersonal communication}. In \bibinfo{booktitle}{\emph{2013 IEEE RO-MAN}}. IEEE, \bibinfo{pages}{226--231}.
\newblock


\bibitem[Sakashita et~al\mbox{.}(2022)]%
        {sakashita2022remotecode}
\bibfield{author}{\bibinfo{person}{Mose Sakashita}, \bibinfo{person}{E~Andy Ricci}, \bibinfo{person}{Jatin Arora}, {and} \bibinfo{person}{Fran{\c{c}}ois Guimbreti{\`e}re}.} \bibinfo{year}{2022}\natexlab{}.
\newblock \showarticletitle{RemoteCoDe: Robotic Embodiment for Enhancing Peripheral Awareness in Remote Collaboration Tasks}.
\newblock \bibinfo{journal}{\emph{Proceedings of the ACM on Human-Computer Interaction}} \bibinfo{volume}{6}, \bibinfo{number}{CSCW1} (\bibinfo{year}{2022}), \bibinfo{pages}{1--22}.
\newblock


\bibitem[Sakashita et~al\mbox{.}(2023)]%
        {sakashita2023remotion}
\bibfield{author}{\bibinfo{person}{Mose Sakashita}, \bibinfo{person}{Ruidong Zhang}, \bibinfo{person}{Xiaoyi Li}, \bibinfo{person}{Hyunju Kim}, \bibinfo{person}{Michael Russo}, \bibinfo{person}{Cheng Zhang}, \bibinfo{person}{Malte~F Jung}, {and} \bibinfo{person}{Fran{\c{c}}ois Guimbreti{\`e}re}.} \bibinfo{year}{2023}\natexlab{}.
\newblock \showarticletitle{ReMotion: Supporting Remote Collaboration in Open Space with Automatic Robotic Embodiment}. In \bibinfo{booktitle}{\emph{Proceedings of the 2023 CHI Conference on Human Factors in Computing Systems}}. \bibinfo{pages}{1--14}.
\newblock


\bibitem[Scott et~al\mbox{.}(2004)]%
        {scott2004territoriality}
\bibfield{author}{\bibinfo{person}{Stacey~D Scott}, \bibinfo{person}{M~Sheelagh~T Carpendale}, {and} \bibinfo{person}{Kori Inkpen}.} \bibinfo{year}{2004}\natexlab{}.
\newblock \showarticletitle{Territoriality in collaborative tabletop workspaces}. In \bibinfo{booktitle}{\emph{Proceedings of the 2004 ACM conference on Computer supported cooperative work}}. \bibinfo{pages}{294--303}.
\newblock


\bibitem[Shrestha et~al\mbox{.}(2018)]%
        {shrestha2018communicating}
\bibfield{author}{\bibinfo{person}{Moondeep~C Shrestha}, \bibinfo{person}{Tomoya Onishi}, \bibinfo{person}{Ayano Kobayashi}, \bibinfo{person}{Mitsuhiro Kamezaki}, {and} \bibinfo{person}{Shigeki Sugano}.} \bibinfo{year}{2018}\natexlab{}.
\newblock \showarticletitle{Communicating directional intent in robot navigation using projection indicators}. In \bibinfo{booktitle}{\emph{2018 27th IEEE International Symposium on Robot and Human Interactive Communication (RO-MAN)}}. IEEE, \bibinfo{pages}{746--751}.
\newblock


\bibitem[Sigitov et~al\mbox{.}(2019)]%
        {sigitov2019task}
\bibfield{author}{\bibinfo{person}{Anton Sigitov}, \bibinfo{person}{Andr{\'e} Hinkenjann}, \bibinfo{person}{Ernst Kruijff}, {and} \bibinfo{person}{Oliver Staadt}.} \bibinfo{year}{2019}\natexlab{}.
\newblock \showarticletitle{Task Dependent Group Coupling and Territorial Behavior on Large Tiled Displays}.
\newblock \bibinfo{journal}{\emph{Frontiers in Robotics and AI}}  \bibinfo{volume}{6} (\bibinfo{year}{2019}), \bibinfo{pages}{128}.
\newblock


\bibitem[Smith(2017)]%
        {smith2017self}
\bibfield{author}{\bibinfo{person}{Joel Smith}.} \bibinfo{year}{2017}\natexlab{}.
\newblock \showarticletitle{Self-consciousness}.
\newblock  (\bibinfo{year}{2017}).
\newblock


\bibitem[Stahl et~al\mbox{.}(2018)]%
        {stahl2018social}
\bibfield{author}{\bibinfo{person}{Christoph Stahl}, \bibinfo{person}{Dimitra Anastasiou}, {and} \bibinfo{person}{Thibaud Latour}.} \bibinfo{year}{2018}\natexlab{}.
\newblock \showarticletitle{Social telepresence robots: The role of gesture for collaboration over a distance}. In \bibinfo{booktitle}{\emph{Proceedings of the 11th PErvasive Technologies Related to Assistive Environments Conference}}. \bibinfo{pages}{409--414}.
\newblock


\bibitem[Tang et~al\mbox{.}({[n.\,d.]})]%
        {tang2017collaboration}
\bibfield{author}{\bibinfo{person}{Anthony Tang}, \bibinfo{person}{Omid Fakourfar}, \bibinfo{person}{Carman Neustaedter}, {and} \bibinfo{person}{Scott Bateman}.} \bibinfo{year}{[n.\,d.]}\natexlab{}.
\newblock \showarticletitle{Collaboration with 360° Videochat: Challenges and Opportunities.}
\newblock


\bibitem[Teo et~al\mbox{.}(2019)]%
        {teo2019mixed}
\bibfield{author}{\bibinfo{person}{Theophilus Teo}, \bibinfo{person}{Louise Lawrence}, \bibinfo{person}{Gun~A Lee}, \bibinfo{person}{Mark Billinghurst}, {and} \bibinfo{person}{Matt Adcock}.} \bibinfo{year}{2019}\natexlab{}.
\newblock \showarticletitle{Mixed reality remote collaboration combining 360 video and 3d reconstruction}. In \bibinfo{booktitle}{\emph{Proceedings of the 2019 CHI conference on human factors in computing systems}}. \bibinfo{pages}{1--14}.
\newblock


\bibitem[Tollmar et~al\mbox{.}(1996)]%
        {tollmar1996supporting}
\bibfield{author}{\bibinfo{person}{Konrad Tollmar}, \bibinfo{person}{Ovidiu Sandor}, {and} \bibinfo{person}{Anna Sch{\"o}mer}.} \bibinfo{year}{1996}\natexlab{}.
\newblock \showarticletitle{Supporting social awareness@ work design and experience}. In \bibinfo{booktitle}{\emph{Proceedings of the 1996 ACM conference on Computer supported cooperative work}}. \bibinfo{pages}{298--307}.
\newblock


\bibitem[Trahanias et~al\mbox{.}(2000)]%
        {trahanias2000tourbot}
\bibfield{author}{\bibinfo{person}{Panos Trahanias}, \bibinfo{person}{Antonis Argyros}, \bibinfo{person}{Dimitris Tsakiris}, \bibinfo{person}{Armin Cremers}, \bibinfo{person}{Dirk Schulz}, \bibinfo{person}{Wolfram Burgard}, \bibinfo{person}{Dirk Haehnel}, \bibinfo{person}{Vassilis Savvaides}, \bibinfo{person}{George Giannoulis}, \bibinfo{person}{Mandy Coliou}, {et~al\mbox{.}}} \bibinfo{year}{2000}\natexlab{}.
\newblock \showarticletitle{Tourbot-interactive museum tele-presence through robotic avatars}. In \bibinfo{booktitle}{\emph{Proc. of the 9th International World Wide Web Conference}}.
\newblock


\bibitem[Wang et~al\mbox{.}(2023)]%
        {wang2023gesture}
\bibfield{author}{\bibinfo{person}{Yibo Wang}, \bibinfo{person}{Chenwei Zhang}, \bibinfo{person}{Heqiao Wang}, \bibinfo{person}{Shuya Lu}, {and} \bibinfo{person}{Ray LC}.} \bibinfo{year}{2023}\natexlab{}.
\newblock \showarticletitle{Gesture-Bot: Design and Evaluation of Simple Gestures of a Do-it-yourself Telepresence Robot for Remote Communication}. In \bibinfo{booktitle}{\emph{Companion of the 2023 ACM/IEEE International Conference on Human-Robot Interaction}}. \bibinfo{pages}{102--106}.
\newblock


\bibitem[Weissker et~al\mbox{.}(2020)]%
        {weissker2020getting}
\bibfield{author}{\bibinfo{person}{Tim Weissker}, \bibinfo{person}{Pauline Bimberg}, {and} \bibinfo{person}{Bernd Froehlich}.} \bibinfo{year}{2020}\natexlab{}.
\newblock \showarticletitle{Getting there together: Group navigation in distributed virtual environments}.
\newblock \bibinfo{journal}{\emph{IEEE transactions on visualization and computer graphics}} \bibinfo{volume}{26}, \bibinfo{number}{5} (\bibinfo{year}{2020}), \bibinfo{pages}{1860--1870}.
\newblock


\bibitem[Weissker and Froehlich(2021)]%
        {weissker2021group}
\bibfield{author}{\bibinfo{person}{Tim Weissker} {and} \bibinfo{person}{Bernd Froehlich}.} \bibinfo{year}{2021}\natexlab{}.
\newblock \showarticletitle{Group navigation for guided tours in distributed virtual environments}.
\newblock \bibinfo{journal}{\emph{IEEE Transactions on Visualization and Computer Graphics}} \bibinfo{volume}{27}, \bibinfo{number}{5} (\bibinfo{year}{2021}), \bibinfo{pages}{2524--2534}.
\newblock


\bibitem[Weissker et~al\mbox{.}(2019)]%
        {weissker2019multi}
\bibfield{author}{\bibinfo{person}{Tim Weissker}, \bibinfo{person}{Alexander Kulik}, {and} \bibinfo{person}{Bernd Froehlich}.} \bibinfo{year}{2019}\natexlab{}.
\newblock \showarticletitle{Multi-ray jumping: comprehensible group navigation for collocated users in immersive virtual reality}. In \bibinfo{booktitle}{\emph{2019 IEEE Conference on Virtual Reality and 3D User Interfaces (VR)}}. IEEE, \bibinfo{pages}{136--144}.
\newblock


\bibitem[Wu et~al\mbox{.}(2023)]%
        {wu2023mr}
\bibfield{author}{\bibinfo{person}{Yudan Wu}, \bibinfo{person}{Shanhe You}, \bibinfo{person}{Zixuan Guo}, \bibinfo{person}{Xiangyang Li}, \bibinfo{person}{Guyue Zhou}, {and} \bibinfo{person}{Jiangtao Gong}.} \bibinfo{year}{2023}\natexlab{}.
\newblock \showarticletitle{MR. Brick: designing a remote mixed-reality educational game system for promoting children’s social \& collaborative skills}. In \bibinfo{booktitle}{\emph{Proceedings of the 2023 CHI Conference on Human Factors in Computing Systems}}. \bibinfo{pages}{1--18}.
\newblock


\bibitem[Yang et~al\mbox{.}(2018)]%
        {yang2018shopping}
\bibfield{author}{\bibinfo{person}{Lillian Yang}, \bibinfo{person}{Brennan Jones}, \bibinfo{person}{Carman Neustaedter}, {and} \bibinfo{person}{Samarth Singhal}.} \bibinfo{year}{2018}\natexlab{}.
\newblock \showarticletitle{Shopping over distance through a telepresence robot}.
\newblock \bibinfo{journal}{\emph{Proceedings of the ACM on Human-Computer Interaction}} \bibinfo{volume}{2}, \bibinfo{number}{CSCW} (\bibinfo{year}{2018}), \bibinfo{pages}{1--18}.
\newblock


\bibitem[Yonezawa and Ueda(2014)]%
        {yonezawa2014representation}
\bibfield{author}{\bibinfo{person}{Ken Yonezawa} {and} \bibinfo{person}{Hirotada Ueda}.} \bibinfo{year}{2014}\natexlab{}.
\newblock \showarticletitle{Representation of gaze, mood, and emotion: movie-watching with telepresence robots}. In \bibinfo{booktitle}{\emph{Proceedings of the second international conference on Human-agent interaction}}. \bibinfo{pages}{261--264}.
\newblock


\bibitem[Young et~al\mbox{.}(2020)]%
        {young2020mobileportation}
\bibfield{author}{\bibinfo{person}{Jacob Young}, \bibinfo{person}{Tobias Langlotz}, \bibinfo{person}{Steven Mills}, {and} \bibinfo{person}{Holger Regenbrecht}.} \bibinfo{year}{2020}\natexlab{}.
\newblock \showarticletitle{Mobileportation: Nomadic telepresence for mobile devices}.
\newblock \bibinfo{journal}{\emph{Proceedings of the ACM on Interactive, Mobile, Wearable and Ubiquitous Technologies}} \bibinfo{volume}{4}, \bibinfo{number}{2} (\bibinfo{year}{2020}), \bibinfo{pages}{1--16}.
\newblock


\bibitem[Zhan et~al\mbox{.}(2023)]%
        {zhan2023enable}
\bibfield{author}{\bibinfo{person}{Lishuang Zhan}, \bibinfo{person}{Yancheng Cao}, \bibinfo{person}{Qitai Chen}, \bibinfo{person}{Haole Guo}, \bibinfo{person}{Jiasi Gao}, \bibinfo{person}{Yiyue Luo}, \bibinfo{person}{Shihui Guo}, \bibinfo{person}{Guyue Zhou}, {and} \bibinfo{person}{Jiangtao Gong}.} \bibinfo{year}{2023}\natexlab{}.
\newblock \showarticletitle{Enable Natural Tactile Interaction for Robot Dog based on Large-format Distributed Flexible Pressure Sensors}. In \bibinfo{booktitle}{\emph{2023 IEEE International Conference on Robotics and Automation (ICRA)}}. IEEE, \bibinfo{pages}{12493--12499}.
\newblock


\bibitem[Zhang et~al\mbox{.}(2023)]%
        {zhang2023follower}
\bibfield{author}{\bibinfo{person}{Yan Zhang}, \bibinfo{person}{Ziang Li}, \bibinfo{person}{Haole Guo}, \bibinfo{person}{Luyao Wang}, \bibinfo{person}{Qihe Chen}, \bibinfo{person}{Wenjie Jiang}, \bibinfo{person}{Mingming Fan}, \bibinfo{person}{Guyue Zhou}, {and} \bibinfo{person}{Jiangtao Gong}.} \bibinfo{year}{2023}\natexlab{}.
\newblock \showarticletitle{" I am the follower, also the boss": Exploring Different Levels of Autonomy and Machine Forms of Guiding Robots for the Visually Impaired}. In \bibinfo{booktitle}{\emph{Proceedings of the 2023 CHI Conference on Human Factors in Computing Systems}}. \bibinfo{pages}{1--22}.
\newblock


\end{thebibliography}

\appendix


\end{document}